\documentclass[%
 reprint,
 amsmath,amssymb,
 aps,
]{revtex4-2}

\usepackage{graphicx}
\usepackage{hyperref}
\usepackage{dcolumn}
\usepackage{bm}
\usepackage{braket}
\usepackage{amsmath}
\usepackage{subcaption}
\usepackage{makecell}
\usepackage[dvipsnames]{xcolor}
\usepackage{comment}
\usepackage{algorithm}
\usepackage{algpseudocode}

\newcommand\fig[1]{Fig.\ref{fig:#1}}

\newcommand\tab[1]{Table~\ref{tab:#1}}

\usepackage{ulem} 
\newcommand{\ama}[1]{\textcolor{red}{[AA: #1]}}

\begin{document}

\preprint{APS/123-QED}

\title{Laser interferometry as a robust neuromorphic platform for machine learning}

\author{Amanuel Anteneh}
\email{asa2rc@virginia.edu}
\affiliation{Department of Physics, University of Virginia, 382 McCormick Rd, Charlottesville, VA 22903, USA}

\author{Kyungeun Kim}
\affiliation{Department of Mathematics, The University of British Columbia, Vancouver, BC Canada}
\author{J. M. Schwarz}
\affiliation{Department of Physics, Syracuse University, Syracuse, NY, USA}
\author{Israel Klich}
\affiliation{Department of Physics, University of Virginia, 382 McCormick Rd, Charlottesville, VA 22903, USA}
\affiliation{Max Planck Institute for the Physics of Complex Systems, 01187 Dresden, Germany}
\author{Olivier Pfister}
\email{olivier.pfister@gmail.com}
\affiliation{Department of Physics, University of Virginia, 382 McCormick Rd, Charlottesville, VA 22903, USA}
\affiliation{Charles L. Brown Department of Electrical and Computer Engineering, University of Virginia, 351 McCormick Road, Charlottesville, VA 22903, USA}

\date{\today}

\begin{abstract}
We present a method for implementing an optical neural network using only linear optical resources, namely field displacement and interferometry applied to coherent states of light. 
The nonlinearity required for learning in a neural network is realized via an encoding of the input into phase shifts allowing for far more straightforward experimental implementation compared to previous proposals for, and demonstrations of, $\textit{in situ}$ inference. 
Beyond $\textit{in situ}$ inference, the method enables $\textit{in situ}$ training by utilizing established techniques like parameter shift methods or physical backpropagation to extract gradients directly from measurements of the linear optical circuit.
We also investigate the effect of photon losses and find the model to be very resilient to these.
\end{abstract}

\maketitle

\section{Introduction}
Modern deep learning models have reached unprecedented sizes with recent large language models having billions of trainable parameters and taking months of time for pretraining on clusters of multiple graphics processing units (GPUs)~\cite{brown2020language, raffel2020exploring, touvron2023llama}. 
The training and inference phases for these models require increasingly large amounts of energy due to the energy-efficiency limitations imposed by the modern silicon-based digital computers on which they are implemented, a phenomenon referred to as the ‘von Neumann bottleneck’. 
This bottleneck is a consequence of the separation of computing and memory into distinct units in the current von-Neuman computer architecture which necessitates costly data transfer between these units during computation~\cite{markovic2020physics, christensen20222022}. 
This challenge has motivated the development of neuromorphic computing, a field that seeks to construct hardware architectures inspired by the remarkable energy efficiency of the human brain~\cite{markovic2020physics, christensen20222022}. 
Consequently, the intersection of machine learning (ML) and neuromorphic computing has lead to advances in the development of physical neural networks (PNNs)~\cite{momeni2025training}. 
These neural networks seek to harness the capabilities of various analog systems to improve both the speed and resource efficiency with which these models can be trained and deployed for inference.  

A promising area being explored for neuromorphic computing, and thus for the implementation of PNNs, is the field of photonics~\cite{christensen20222022, momeni2025training}. 
Neuromorphic computing implementations based on photonics have several advantages such as higher energy efficiency, throughput, speed and parallelization capability~\cite{christensen20222022}. 
However, a limitation of current approaches based on photonics is scalability and difficulty of implementing low power optical nonlinearities~\cite{christensen20222022}. 
On the scalability front quantum photonics, particularly with continuous variable (CV) cluster states, can provide a remedy as much work has been done on scaling these systems to large sizes for measurement-based quantum computing schemes~\cite{Chen2014,Yoshikawa2016,Larsen2019,Asavanant2019}.
The original proposal for a CV quantum neural  network~\cite{killoran2019continuous}, as well as derivative work~\cite{austin2025hybrid}, utilized optical Kerr nonlinearities which are difficult to implement experimentally. An alternative proposal utilized repeat-until-success photon number resolving measurements on ancillary modes to introduce nonlinearity into the network~\cite{bangar2023experimentally}. However, the implementation of quantum optical nonlinearities is still a challenging task as in the classical optics case. 
One solution that has been explored in the classical optics approach is to utilize hybrid architectures based on optical-digital conversion~\cite{shen2017deep, wang2022optical, pai2023experimentally} 
or opto-electronic conversion \cite{shainline2017superconducting, williamson2019reprogrammable, hamerly2019large, ashtiani2022chip, chen2023deep, xue2024fully, Bandyopadhyay2024}. 
Another approach uses the intrinsic nonlinearity within coupled semiconductor nanolaser arrays~\cite{ji2025photonic}. 

However, learning nonlinearity should not be equated with physics nonlinearity. 
Indeed, recent breakthroughs in optical implementations~\cite{PT2024,wanjura2024fully,yildirim2024nonlinear, xia2024nonlinear} stemmed from the realization that optical nonlinearities in the field variables are not needed to implement learning~\cite{PT2024} if one doesn't encode the data in them exclusively and, rather, uses the experimental parameters for that purpose: a simple optical phase shift $\theta$ imparts field transformations $\cos\theta, \sin\theta$ that are nonlinear in $\theta$ (though linear in the fields). 

This has recently been proposed~\cite{wanjura2024fully} and demonstrated~\cite{yildirim2024nonlinear,xia2024nonlinear} using classical linear wave scattering.  

\begin{figure*}[ht]
    \centering    \includegraphics[width=0.6\textwidth]{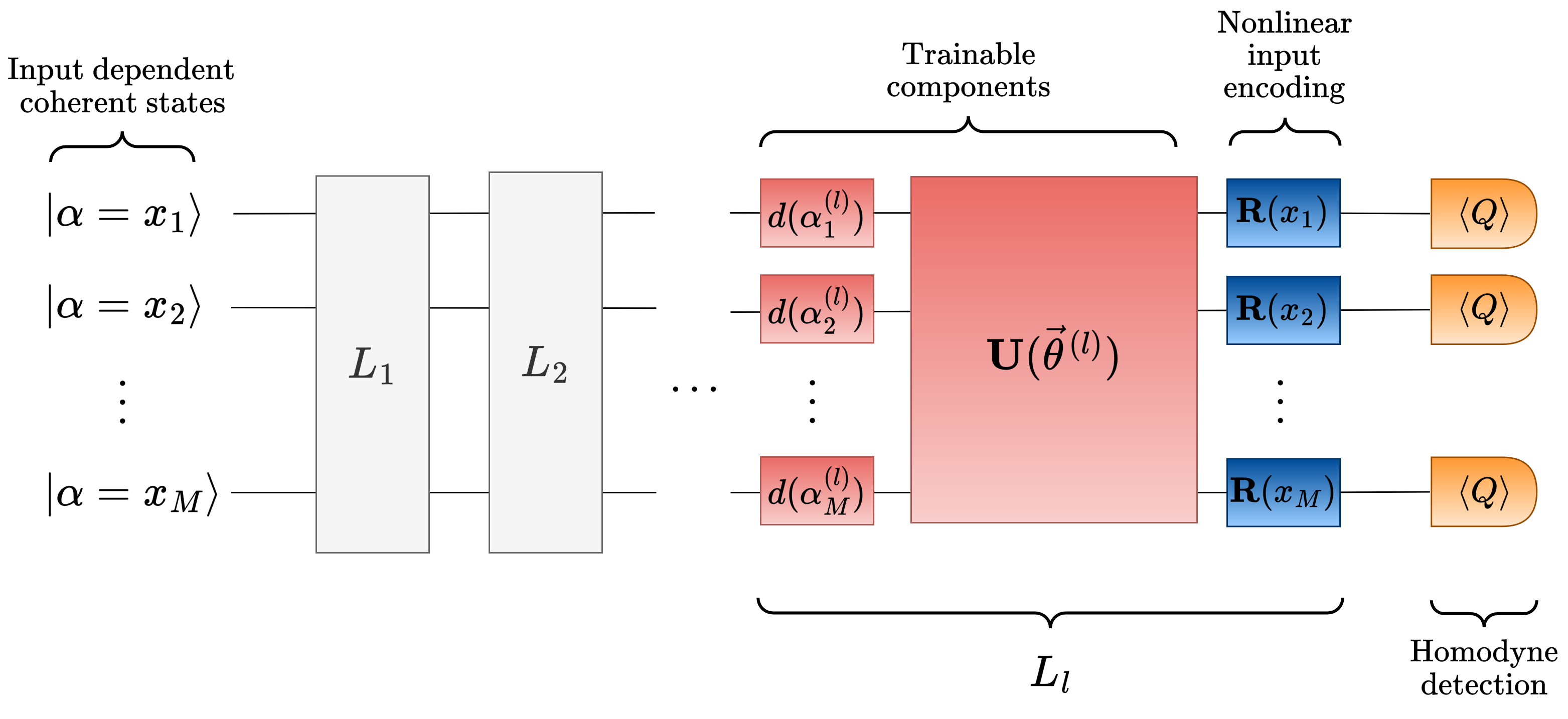}
    \caption{Schematic diagram of linear optical circuit for nonlinear processing with an input $\vec{x} = [x_1,x_2,...,x_M]^T \in \mathbb{R}^M$. The nonlinearity is realized by encoding the input into quadrature phase shifts $\mathbf R(x)$ while affine transformations are carried out using an $M \times M$ interferometer $\mathbf U(\vec{\theta})$ and displacements $d(\alpha)$ with the model output being extracted using homodyne measurements of the $Q$ quadrature. Note that $\vec{\theta}$ denotes the set of transmittivity angles for the $M(M-1)/2$ beamsplitters that comprise the interferometer. }
    \label{fig:optical_linear_nn}
\end{figure*}

In this paper, we propose a different linear optics encoding: a multimode laser interferometer with added amplitude and phase displacements, see \fig{optical_linear_nn}. This approach is considerably simpler than the previous ones in terms of experimental implementation, does not utilize digital linear layers for postprocessing, and is also compatible with the current state of integrated photonics~\cite{Bogaerts2020}.

It is interesting---in particular to set the stage for future extensions to the quantum regime---to deliberately adopt a quantum optical description of the optical circuit, even though it sits firmly within the classical border in this paper. Indeed, three classes of optical circuits can be defined: {\it(i)}, classical linear optics (laser interferometers with photodetectors of dark current well above the single-photon level), {\it(ii)}, second-order nonlinear optics yielding only two-wave mixing (quantum squeezers, e.g.\ undepleted-pump optical parametric oscillators) and, {\it(iii)}, true three-wave mixing as well as third- and higher order nonlinear optics. While {\it(ii)} and {\it(iii)} both enter the realm of quantum optics, only {\it(iii)} involves nonlinear quantum evolution, a.k.a.\ non-Gaussian Wigner function gates and/or states of the quantum fields. This is why such operations---be they implemented via third-order nonlinearities or Fock state projection---were included in previous works on quantum neural networks~\cite{killoran2019continuous,bangar2023experimentally,austin2025hybrid}. Again, the gist of linear optics implementations~\cite{PT2024,wanjura2024fully,yildirim2024nonlinear,xia2024nonlinear} is to eschew relying solely on field encoding, and hence the need for optical nonlinearities, in favor of optical parameter encoding. 

\section{Laser interferometer neural network}

\subsection{Gaussian quantum formalism}

As mentioned above, we adopt here the more general quantum optical formalism of qumodes in continuous-variable quantum information~\cite{Braunstein2005a,Weedbrook2012,Pfister2019}, for both the sake of generality and in anticipation of future extensions beyond the classical context of this work.
Consider an $M$-qumode system~\cite{adesso2014continuous} given by the vector of quadrature operators $Q_j=(a_j+a_j^\dag)/\sqrt2$ and $P_j=i(a_j^\dag-a_j)/\sqrt2$ of each qumode $j$, $a_j$ being the photon annihilation operator,
\begin{align}
    \vec{R} = \begin{pmatrix}
    Q_1, P_1,\cdots, Q_M,P_M
    \end{pmatrix}^T
\end{align}
the vector of quantum expectation values $\vec{r} \in \mathbb{R}^{2M}$ is defined as
\begin{align}
    \vec{r}
    = \langle \vec{R}
    \rangle,
\end{align}
and the covariance matrix $\mathbf\Sigma \in \mathbb{R}^{2M \times 2M}$ is given by
\begin{align}
    \Sigma_{ij} = \langle \vec{R}_i \vec{R}_j + \vec{R}_j \vec{R}_i \rangle - 2 \langle\vec{R}_i\rangle \langle\vec{R}_j\rangle.
\end{align}
These two mathematical objects provide a complete description of the quantum system in the Gaussian Wigner function case for all gates and states but only the first moments will be relevant for our work. 
The unitary evolution of the system under Gaussian unitaries is given by a symplectic matrix $\mathbf S$ which acts on the vector of means and covariance matrix as~\cite{adesso2014continuous}
\begin{align}
    \vec{r} &\mapsto \mathbf S \vec{r} \\
    \mathbf \Sigma &\mapsto \mathbf S \mathbf \Sigma \mathbf S^T.
\end{align}
The symplectic matrix representations of the single-mode phase shift $\mathbf R(\phi)$ and two-mode beamsplitter $\mathbf B(\theta)$ unitaries are respectively given by~\cite{Weedbrook2012}
\begin{align}
    \mathbf R(\phi) = 
    \begin{pmatrix} 
    \cos\phi & \sin\phi \\
    -\sin\phi  &  \cos\phi
    \end{pmatrix}
\end{align}
and
\begin{align}
    \mathbf B(\theta) = 
    \begin{pmatrix} 
    \cos\theta\times \mathbb{I}_2 & \sin\theta\times\mathbb{I}_2 \\
    -\sin\theta\times\mathbb{I}_2  &  \cos\theta\times\mathbb{I}_2
    \end{pmatrix}
\end{align}
 where $\mathbb{I}_2$ is the $2\times 2$ identity matrix.
The special case of displacements by $M$ complex amplitudes $\vec{\alpha}=(\alpha_1,...,\alpha_M)^T$ is described by a displacement vector 
\begin{align}
    \vec d(\vec{\alpha}) = \sqrt{2} \begin{pmatrix}
    \textrm{Re}[\vec{\alpha}_1], \textrm{Im}[\vec{\alpha}_1], \dots, \textrm{Re}[\vec{\alpha}_M], \textrm{Im}[\vec{\alpha}_M] 
    \end{pmatrix}^T
\end{align}
which transforms the vector of means as
\begin{align}
    \vec{r} \mapsto \vec{r} + \vec d(\vec{\alpha}).
\end{align}
Finally, note that all quantum states of light at the input and throughout the circuit (\fig{optical_linear_nn}) will be coherent states.
We simulate the circuit using the \texttt{PyTorch} framework \cite{paszke2019pytorch}.

\begin{figure*}[t]
\centering
\begin{minipage}{0.95\textwidth}
\begin{algorithm}[H]
\caption{ONN training loop with Adam algorithm}
\label{alg:adam}
\begin{algorithmic}[1]
\Require Initial parameters $\theta_0$, 
initial learning rate $\gamma_0$, 
maximum number of epochs $T$, exponential decay rates $\beta_1, \beta_2 \in [0,1)$, 
batch size $B$, 
dataset $D = \{(x^{(1)}, y_{\textrm{tar}}^{(1)}), \dots, (x^{(k)}, y_{\textrm{tar}}^{(k)})\}$ of $k$ labeled training examples,
learning rate decay rate $\chi$,
small constant $\epsilon$ for numerical stability (typically set to $10^{-8}$) 
\State $m_0 \leftarrow 0$ \Comment{Initialize first moment variable}
\State $v_0 \leftarrow 0$ \Comment{Initialize second moment variable}
\State $t \leftarrow 0$ \Comment{Current epoch number}
\While{$t< T$}
    \State $t \leftarrow t + 1$
    \While{all data points in $D$ not yet sampled} \Comment{An epoch is one pass over all data points}
    \State Sample random batch of $B$ data points, $D_B=\{(x^{(1)}, y_{\textrm{tar}}^{(1)}), \dots, (x^{(B)}, y_{\textrm{tar}}^{(B)})\}$, without replacement from $D$
    \State Compute ONN predictions, $y=\{f(x^{(1)};\theta_{t-1}), \dots, f(x^{(B)};\theta_{t-1})\}$, from optical measurements 
    \State $g_t \leftarrow \frac{1}{B}\sum_{i}\nabla_{\theta_{t-1}} \mathcal{L}(y^{(i)}, y_{\textrm{tar}}^{(i)})$ \Comment{Compute average Euclidean gradient over batch}
    \State $m_t \leftarrow \beta_1 m_{t-1} + (1-\beta_1) g_t$ \Comment{Update biased first moment estimate}
    \State $v_t \leftarrow \beta_2 v_{t-1} + (1-\beta_2) g_t \odot g_t$ \Comment{Update biased second moment estimate. $\odot$ denotes Hadamard product.}
    \State $\hat{m}_t \leftarrow \dfrac{m_t}{1-\beta_1^t}$ \Comment{Correct bias in first moment estimate}
    \State $\hat{v}_t \leftarrow \dfrac{v_t}{1-\beta_2^t}$ \Comment{Correct bias in second moment estimate}
    \State $\theta_t \leftarrow \theta_{t-1} - 
        \gamma \dfrac{\hat{m}_t}{\sqrt{\hat{v}_t} + \epsilon}$ \Comment{Compute updated optical circuit parameters}
    \EndWhile
    \State $\gamma_t \leftarrow \gamma_{0}\chi^{t}$ \Comment{Decay learning rate on exponential schedule}
\EndWhile \\
\Return $\theta_t$
\end{algorithmic}
\end{algorithm}

\end{minipage}
\end{figure*}

\subsection{Nonlinear learning with linear optics}
The circuit architecture is shown in \fig{optical_linear_nn}.
The inputs to the circuit are encoded both in the input coherent state amplitudes and, crucially, in the parameters $\phi$ of phase shift operations $\mathbf R(\phi)$. 
The latter enable the output of our optical neural network (ONN), which we define as the expectation values $q_j=\langle Q_j\rangle$, measured using homodyne detection, to become a nonlinear function of the input parameters since the symplectic matrix defining the phase shift operation contains nonlinear functions of the phase angle $\phi$. 
Compared to many optical nonlinearities this approach is significantly more energy efficient. 
Note that we use the expectation value of the quadrature as our output as these can take on the value of any real number similar to the output of a classical neural network.

To approximate a fully connected linear layer present in a classical neural network we use an $M \times M$ interferometer containing beamsplitters of tunable transmissivity angle $\theta$ and fixed zero phase. 
To maintain consistency with classical neural networks' fully connected layers, we employ beamsplitter operations between all $M(M-1)/2$ pairs of modes, contrasting with conventional nearest-neighbor Mach-Zehnder interferometer (MZI) approaches~\cite{reck1994experimental, Clements2016}.  
Specifically, given the mode indices $\{1,2,...,M\}$ the interferometer applies a beamsplitter operation between each pair of modes in the set of all two-element subsets denoted as
\begin{align}
    \binom{\{1,2,\dots,M\}}{2}.
\end{align}
The bias units from a classical neural network are realized by the displacements applied prior to the interferometer in Fig.\ref{fig:optical_linear_nn}.

Note that since the interferometer is represented by a unitary matrix we can only apply unitary transformation of the optical fields but using unitary matrices over general linear layers has been shown to be beneficial in avoiding problems related to vanishing and exploding gradients during training of very deep or recurrent neural networks~\cite{jing2017tunable}.
However, if further expressivity is needed one could achieve this while keeping energy consumption minimal by applying tunable optical attenuators between two interferometers~\cite{shen2017deep}. 
Note also that tunable phase shift and squeezing operations could also be added but we forgo their use for two reasons: {\it(i)}, we aim to minimize the number of tunable parameters in an attempt to better gauge model expressivity and, {\it(ii)}, in a related way, we seek to first evaluate the full measure of classical resources before turning to quantum ones---which may or may not yield a learning advantage---in future work. 

\section{Training}

\begin{figure*}[ht]
    \centerline{\hfill
    \centering
    \begin{subfigure}[h]{0.39\textwidth}
        \centering
        \includegraphics[width=\textwidth]{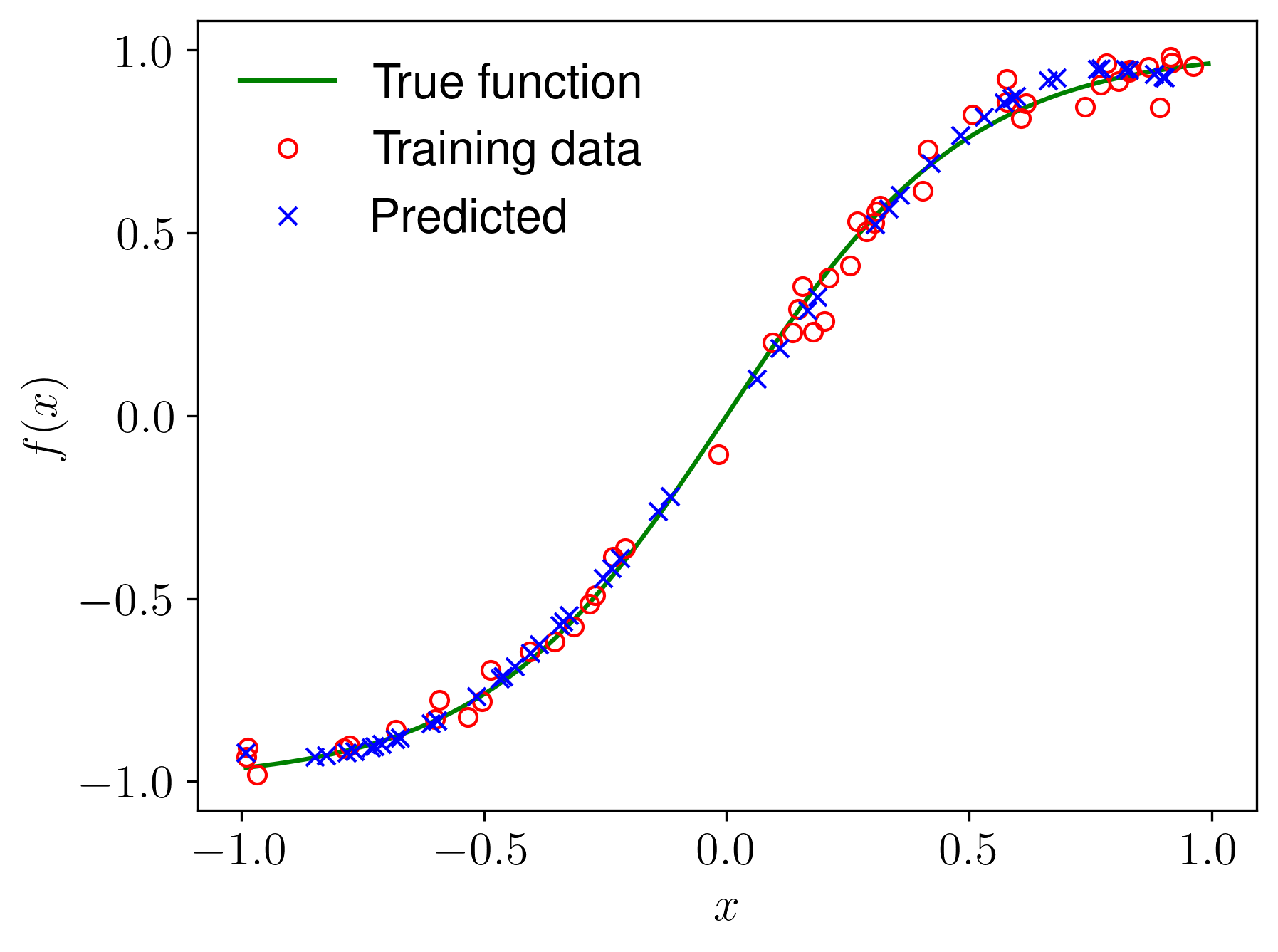}
        \caption{Curve fit for $\tanh(2x)$.}
        \label{fig:reg1}
    \end{subfigure}
    \hfill
    \begin{subfigure}[h]{0.39\textwidth}
        \centering
        \includegraphics[width=\textwidth]{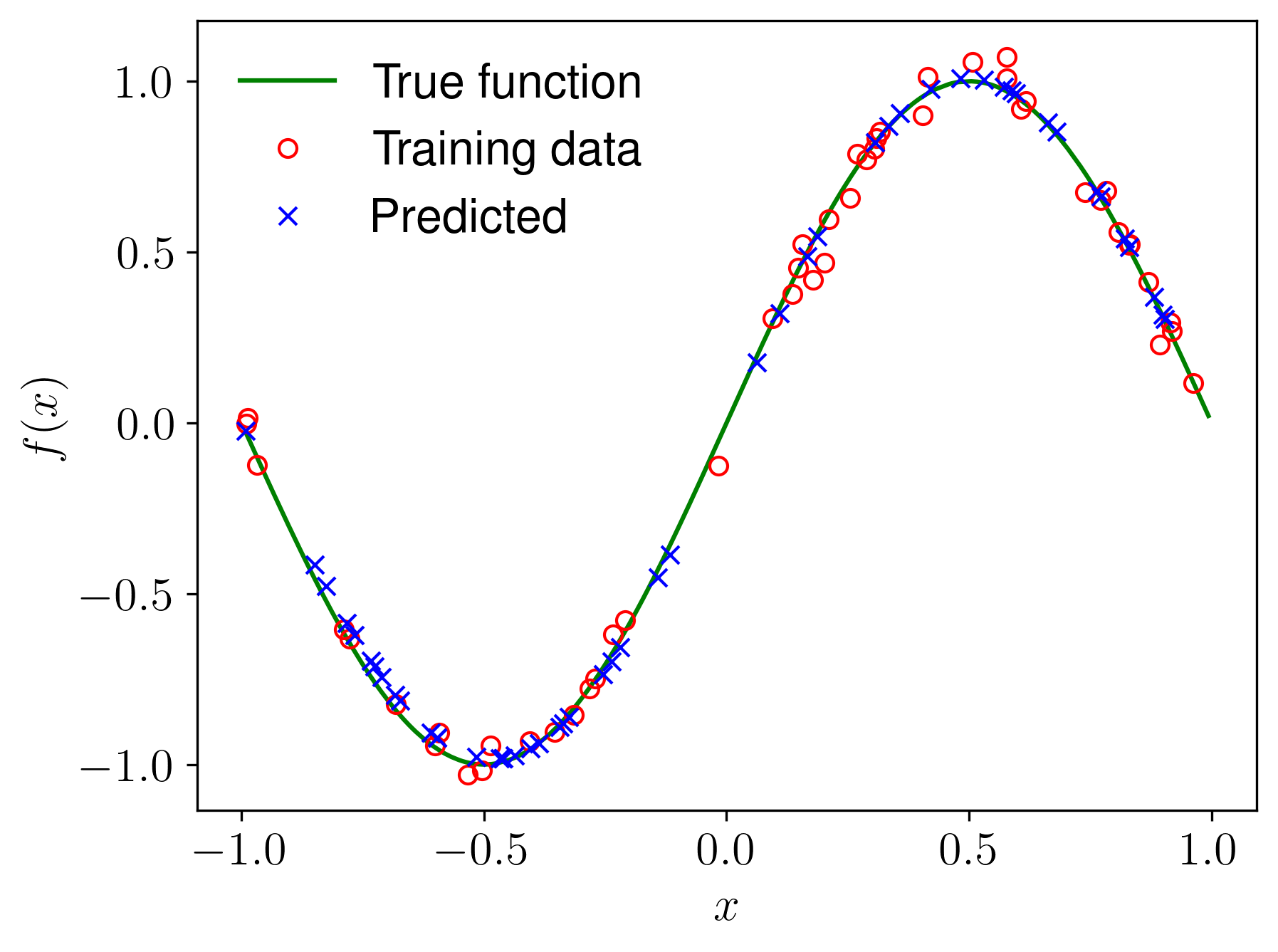}
        \caption{Curve fit for $\sin(\pi x)$.}
        \label{fig:reg3}
    \end{subfigure}
    \hfill}
    {\color{white}space}
    \centerline{\hfill
    \begin{subfigure}[hb]{0.39\textwidth}
        \centering
    \includegraphics[width=\textwidth]{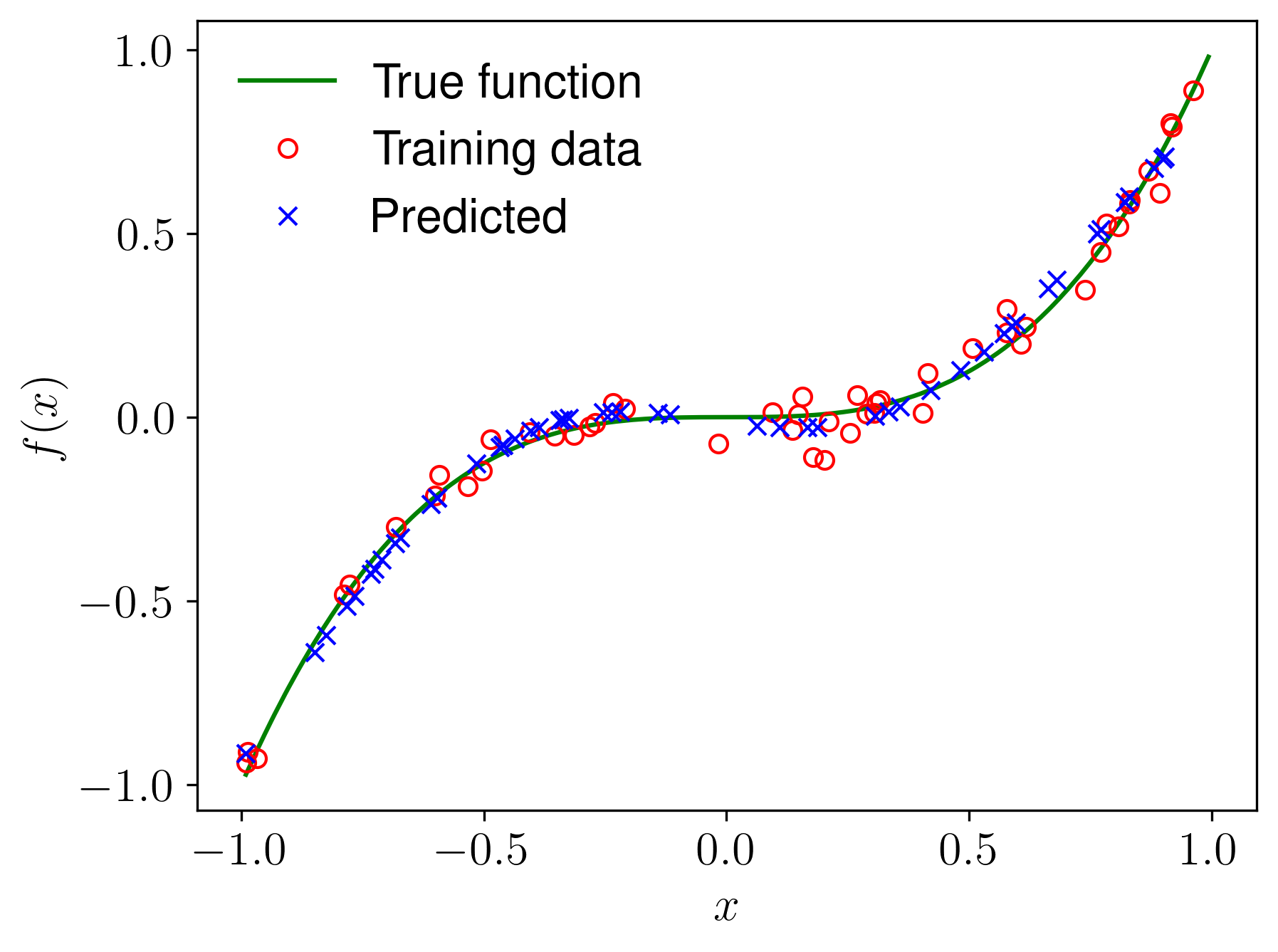}
        \caption{Curve fit for $x^3$.}
        \label{fig:reg3}
    \end{subfigure}
    \hfill
    \begin{subfigure}[hb]{0.39\textwidth}
        \centering
    \includegraphics[width=\textwidth]{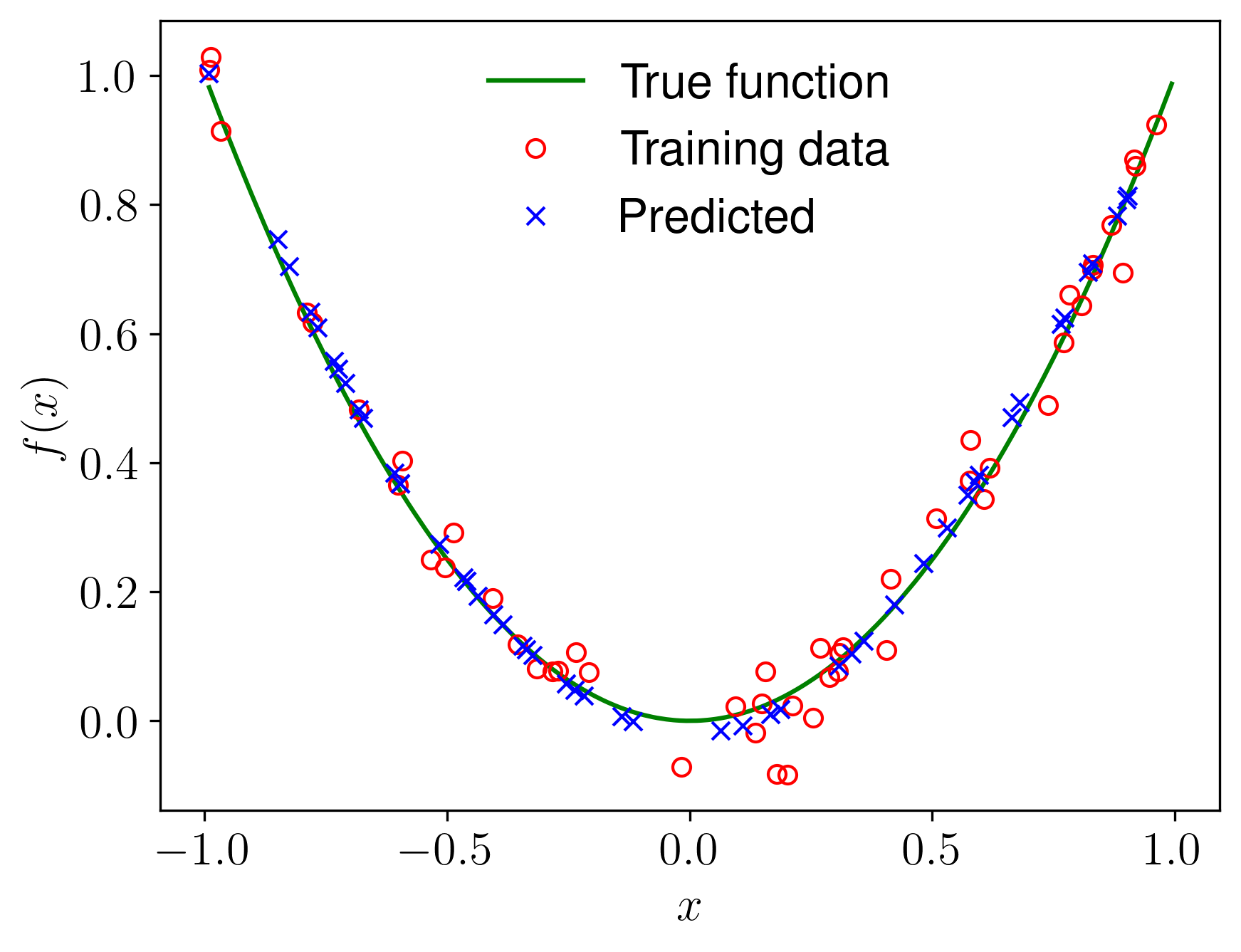}
        \caption{Curve fit for $x^2$.}
        \label{fig:reg4}
    \end{subfigure}
    \hfill}
    \caption{Regression results with Gaussian CV neural network. We set the standard deviation of the Gaussian noise to $\sigma=0.05$. The noisy training data are shown as red circles. The output prediction for test set values of $x$ are shown as blue crosses.}
    \label{fig:lossless_regression}
\end{figure*}

\subsection{Gradient descent}

Neural networks, like many other ML models, are trained by minimize an error function, called the loss function~\cite{bishop2023deep}. 
For supervised learning tasks this function is computed using the networks output $y=f(x;\theta)$, where $f(x;\theta)$, i.e. the model, denotes a function of $x$ parameterized by $\theta$ and a ground truth label which constitutes the correct output $y_{\text{tar}}$. 
The algorithm of choice for the training of neural networks is the stochastic gradient descent algorithm~\cite{bishop2023deep}.
In it's simplest form the algorithm minimizes the loss function $\mathcal{L}$ using the following update rule on the model parameters $\theta$
\begin{align}
    \theta \mapsto \theta - \gamma \nabla_\theta \mathcal{L}
\end{align}
where $\gamma$ is the learning rate and $\nabla_\theta \mathcal{L}$ is the gradient of the loss function with respect to the model parameters $\theta$.
For example for the squared error loss function 
\begin{align}
\mathcal{L}(y, y_{\textrm{tar}})=(y_{\text{tar}} - y)^2=(y_{\text{tar}} - f(x;\theta))^2
\end{align}
using the chain rule formula 
\begin{align}
    \frac{\partial \mathcal{L}}{\partial \theta_i} 
    =\frac{\partial \mathcal{L}}{ \partial y}\ \frac{\partial y}{ \partial \theta_i}
\end{align}
we have
\begin{align}
    \frac{\partial \mathcal{L}}{\partial \theta_i} = 2(y - y_{\text{tar}})  \ \frac{\partial y}{ \partial \theta_i}
\end{align}
where
\begin{align}
    \frac{\partial y}{ \partial \theta_i}
    = \frac{\partial f(x;\theta)}{ \partial \theta_i}
    = \frac{\partial  q}{ \partial \theta_i}
\end{align}
is the gradient of the expectation value of the measured observable with respect to the circuit parameters.
Typically \textit{in silico} training of classical and quantum neural networks utilize the backpropagation algorithm to compute these gradients using reverse-mode automatic differentiation~\cite{bishop2023deep}.

\subsection{Feasibility of \textit{in situ} training}
To perform training \textit{in situ} one will need a way of computing the gradients of the loss function with respect to network parameters using the physical system itself~\cite{momeni2025training}. 
To this end many methods have been proposed for extracting gradients of PNN parameters such as equilibrium propagation for energy based models~\cite{scellier2017equilibrium, scellier2024quantum, wanjura2025quantum},  scatter backpropagation for driven-dissipative  nonlinear coupled-mode systems~\cite{cin2025training} and Hamiltonian echo backpropagation for lossless time-reversible Hamiltonian systems \cite{lopez2023self}. 
Here we present two well known methods for computing gradients \textit{in situ} which are applicable to our optical circuit.

\subsubsection{Parameter-shift method}
For quantum circuits since our output is an expectation value of an observable, e.g. in our case one of the field quadratures, these gradients are also expectation values.
The purely Gaussian nature of our circuit makes \textit{in situ} training much more feasible than previous proposals~\cite{killoran2019continuous,bangar2023experimentally,austin2025hybrid} which use arbitrary numbers of non-Gaussian operations. 
For any purely Gaussian quantum circuit $U(\vec\theta)$, $\vec\theta=(\theta_i)_i^T$, it has been proven that the gradient elements $\partial_{\theta_i} \langle U^\dag(\vec\theta) O U(\vec\theta)\rangle$  of the expectation values of the output observable $O$ with respect to circuit parameters $\theta_i$ can be computed via a simple shift rule of parameter $\theta_i$~\cite{schuld2019evaluating}. 
Here, $O$ must be a low-degree polynomial in the field quadratures. 
Remarkably, these gradient elements can be obtained by using the same circuit that is used for inference, the only modification being that the value of $\theta_i$ is shifted by any nonzero constant twice. 
This result significantly simplifies the process of computing gradients for \textit{in situ} training of Gaussian quantum circuits. 

As a concrete example for the beamsplitter gate consider a circuit which applies a Gaussian unitary $G_1$ followed by a zero phase beamsplitter $BS(\theta)$ and another Gaussian unitary $G_2$ to a two-mode vacuum state input. 
The expectation value of the $Q$ quadrature is given by 
\begin{align}
     q  = \bra{\psi_\theta}Q\ket{\psi_\theta},
\end{align}
where $\ket{\psi_\theta}=G_2BS(\theta)G_1\ket{0}$.
The gradient of this expectation value with respect to the beamsplitter parameter $\theta$ is given by the following parameter shift rule~\cite{schuld2019evaluating}
\begin{align}
    \frac{\partial  q}{\partial \theta} = \frac{1}{2}
    \left[
    \bra{\psi_{\theta+\frac{\pi}{2}}}Q\ket{\psi_{\theta+\frac{\pi}{2}}} - \bra{\psi_{\theta-\frac{\pi}{2}}}Q\ket{\psi_{\theta-\frac{\pi}{2}}} 
    \right].
\end{align}
Similarly in the case where the beamsplitter operation is replaced with a displacement operation $D(\alpha)$, where $\alpha=re^{i\phi}$, the parameter shift rules are given by
\begin{align}
    \frac{\partial q}{\partial r} &= \frac{1}{2s}
    \left[
    \bra{\psi_{r+s}}Q\ket{\psi_{r+s}} - \bra{\psi_{r-s}}Q\ket{\psi_{r-s}} 
    \right],
    \\ 
    \frac{\partial q}{\partial \phi} &= \frac{1}{2}
    \left[
    \bra{\psi_{\phi+\frac{\pi}{2}}}Q\ket{\psi_{\phi+\frac{\pi}{2}}} - \bra{\psi_{\phi-\frac{\pi}{2}}}Q\ket{\psi_{\phi-\frac{\pi}{2}}} 
    \right],
\end{align}
where $s$ is any nonzero real number.
Similar rules are known for the squeezing and phase shift operations.
These parameter shift rules, which are currently only known for Gaussian gates~\cite{bergholm2018pennylane}---which is of course the case here---provide \textit{exact} expressions for the gradients and thus provide an unbiased estimator of the partial derivative unlike methods based on finite differences which are only exact in the case of infinitesimal shifts~\cite{schuld2019evaluating, sweke2020stochastic}. 
It has also been shown empirically in qubit based systems that the variance in the gradient estimate from using a finite number of shots still allows for good model performance even in the extreme case of only a few shots and in fact these estimators can be used to implement a form of gradient descent for which convergence guarantees have been proven~\cite{sweke2020stochastic}. 
The number of shots used to estimate the gradients can also be tuned as a hyperparameter and algorithms which select the number of shots in an adaptive and frugal manner while still achieving good performance have been developed~\cite{kubler2020adaptive}.
The use of only Gaussian resources  also makes a sample complexity analysis for estimating quadrature means straightforward.
When using the the sample mean $\bar{q}$ to estimate the expectation value of the $Q$ quadrature the variance of the estimator is given by~\cite{devore2021modern}
\begin{align}
    \textrm{Var}(\bar{q}) = \frac{\sigma^2}{n},
\end{align}
where $n$ is the number of quadrature samples or shots and $\sigma^2$ is the variance of the quadrature which we know is equal to 1/2 since no squeezing is applied in the circuit and all inputs are coherent states. 
As the sample mean is an unbiased estimator of the population mean the variance is also equal to the mean squared error (MSE) and thus if one wants to estimate the $Q$ quadrature mean to within an MSE of $\epsilon$ one needs
\begin{align}
    n = \frac{\sigma^2}{\epsilon}
\end{align}
quadrature samples.


\subsubsection{Simultaneous perturbation stochastic approximation}
While the parameter shift rule given above provides a simple and unbiased estimator of the necessary gradients it's computational complexity scales linearly with the number of parameters in the model.
There do, however, exist methods for approximating the gradient that are based on similar principles of perturbing the function parameters but that perturb all parameters \textit{simultaneously}.
One example of such a method is simultaneous perturbation stochastic approximation (SPSA)~\cite{spall1998overview}.
This method was used to train the ONNs in \cite{Bandyopadhyay2024, wu2025scaling, xu2026chip}.

The procedure consists of generating a random perturbation vector $\Delta \in \{+\epsilon, -\epsilon\}^{N}$, i.e. a vector of length $N$, where $N$ is the number of parameters in the model, whose components are independent and randomly distributed over the set $\{+\epsilon, -\epsilon\}$. 
Here $\epsilon$ is a hyperparameter that can be optimized and the distribution one uses to generate the values of $\Delta$ must satisfy certain conditions. The Bernoulli $\pm 1$ distribution with $p=0.5$, equivalent to the Rademacher distribution, is an example of such a distribution~\cite{spall1998overview}.
One may then compute 
\begin{align}
        \widehat{\frac{\partial f}{\partial \theta_i}} = \frac{f(x;\Theta+\Delta) - f(x;\Theta-\Delta)}{2c\Delta_i} 
\end{align}
where we use $\widehat{\frac{\partial f}{\partial \theta_i}}$ as this is a stochastic estimate of the gradient and $c$ is a small positive number which typically decreases as the optimization proceeds~\cite{spall1998overview}.
Under reasonably general conditions it can be proven this method and finite differences reach the same level of statistical accuracy given the same number of iterations even though SPSD is $N$ times more efficient and noisy~\cite{spall1998overview}. 

\begin{figure*}[htbp] 
  \centering
  \begin{subfigure}[t]{0.9\textwidth}
    \centering
    \begin{minipage}[b]{0.49\linewidth}
      \centering
      \includegraphics[width=\linewidth]{img/reg/cube_noise_0.05_loss_0.0.png}
    \end{minipage}\hfill
    \begin{minipage}[b]{0.4\linewidth}
      \centering
      \includegraphics[width=\linewidth]{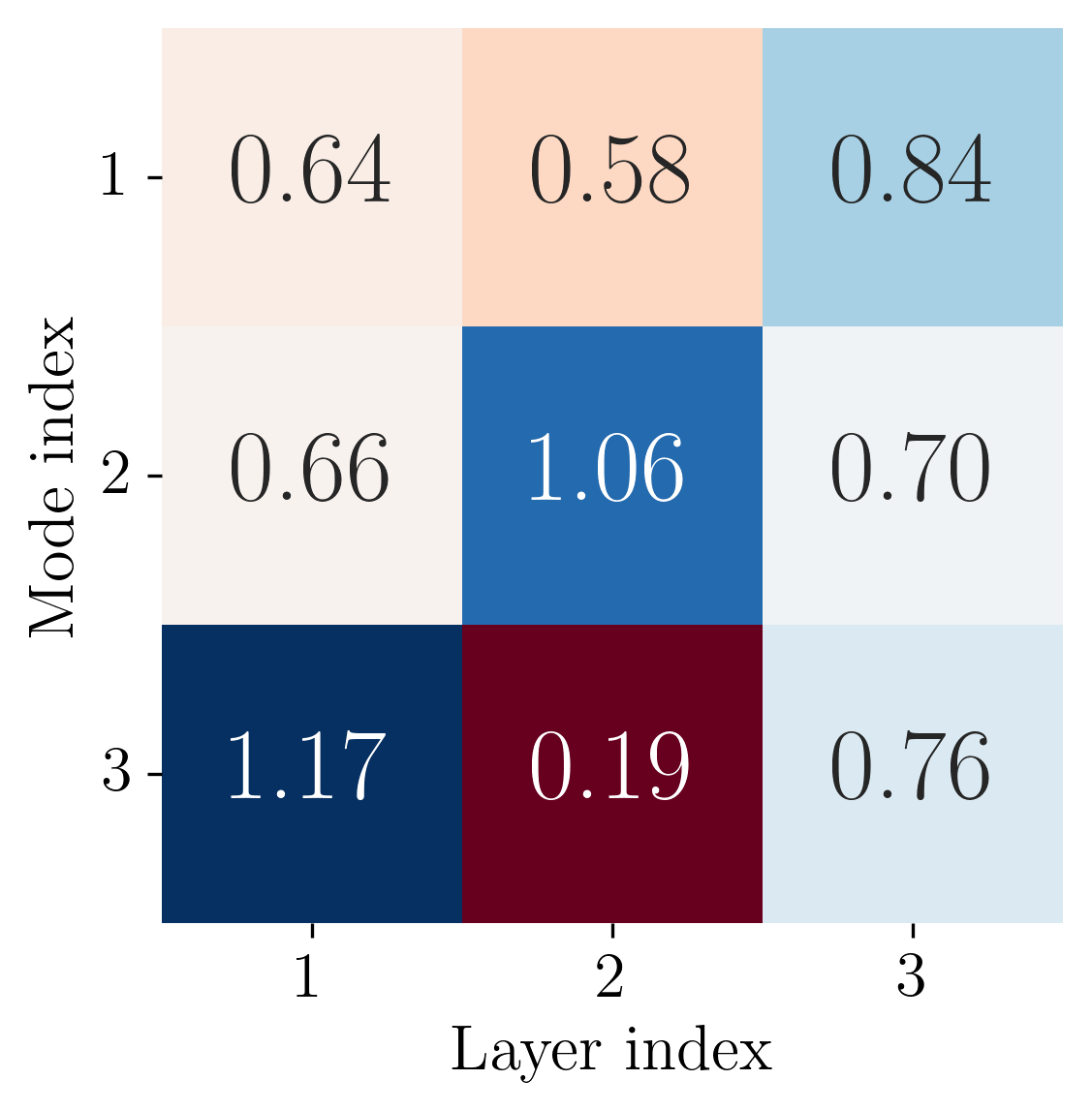}
    \end{minipage}
    \caption{Results with $\eta=0$.}
    \label{fig:reg_lossy_sub1}
  \end{subfigure}\hfill
  \begin{subfigure}[t]{0.9\textwidth}
    \centering
    \begin{minipage}[b]{0.49\linewidth}
      \centering
      \includegraphics[width=\linewidth]{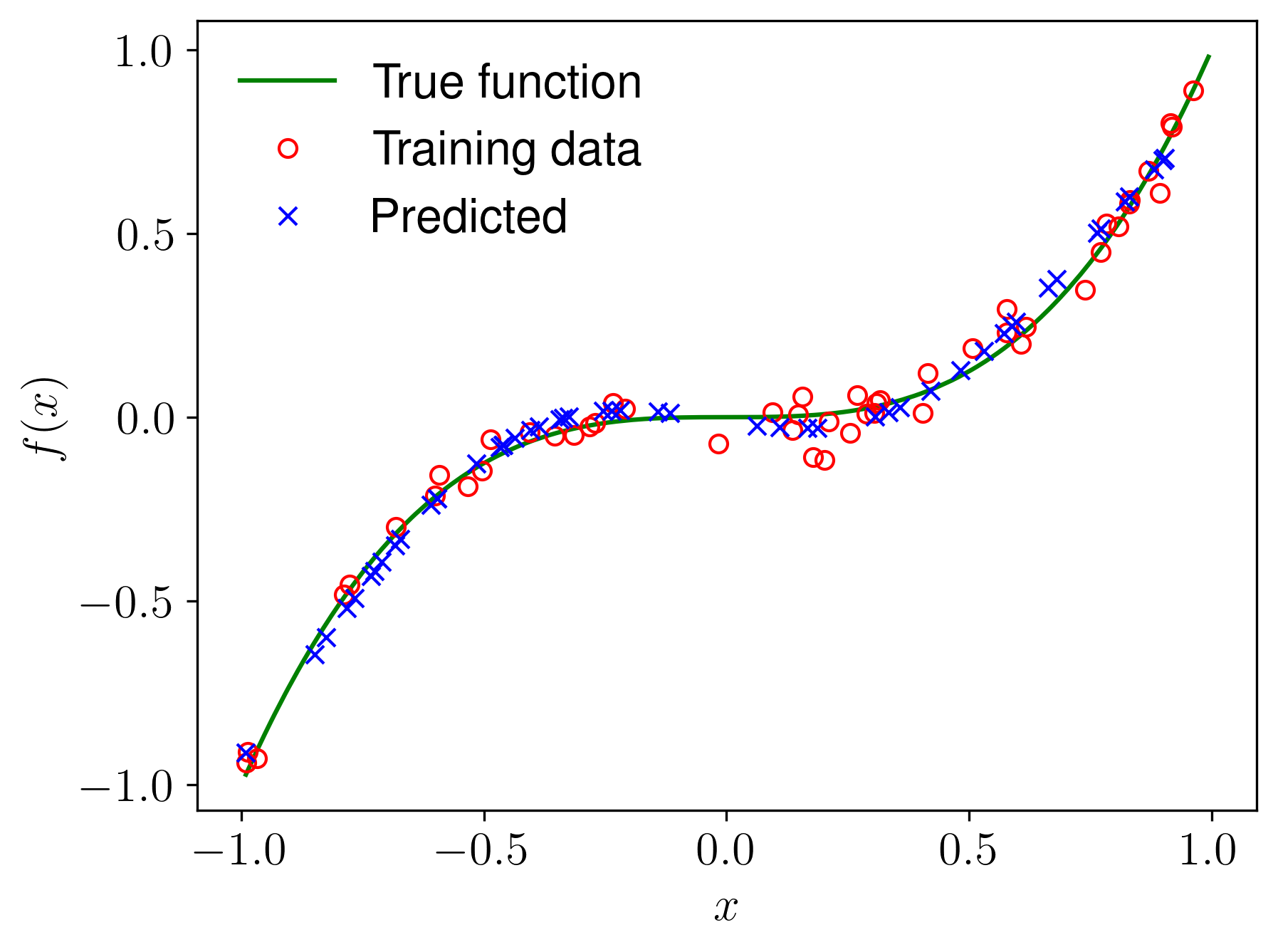}
    \end{minipage}\hfill
    \begin{minipage}[b]{0.4\linewidth}
      \centering
      \includegraphics[width=\linewidth]{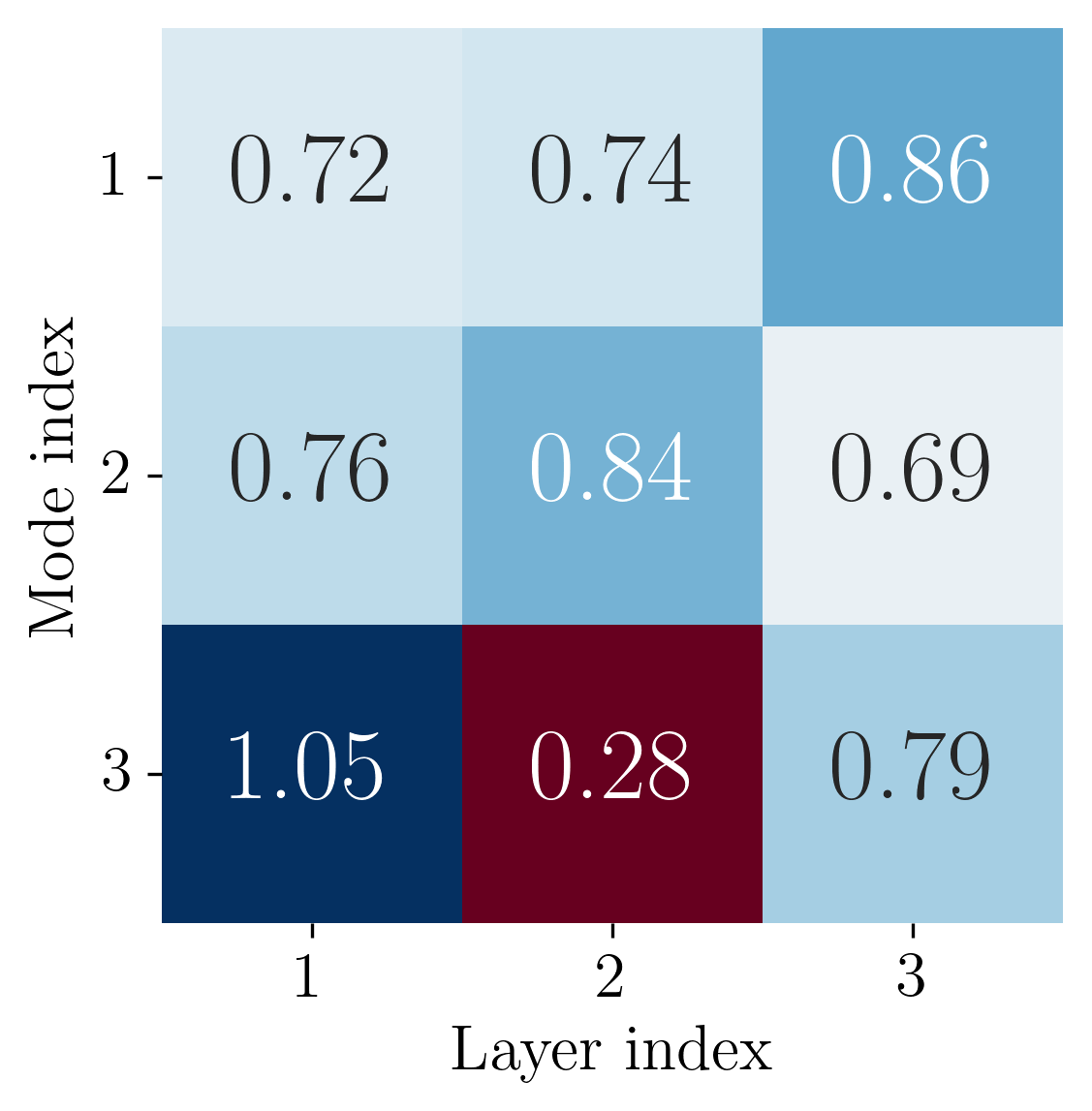}
    \end{minipage}
    \caption{Results with $\eta=0.5$.}
    \label{fig:reg_lossy_sub3}
  \end{subfigure}

  \caption{Effect of photon loss after each layer of the optical circuit when fitting on noisy data from the function $f(x)=x^3$. The heatmap on the right shows the magnitude $|\alpha|$ of the displacements applied to each mode at each layer of the optical circuit.}
  \label{fig:reg_lossy}
\end{figure*}

\section{Numerical Experiments}{\label{num_exp}}

Here we present results from numerical simulations of training our linear optical circuit to solve supervised regression and classification tasks using stochastic gradient descent. 
We utilized the Adam optimizer~\cite{kingma2015adam} implemented in \texttt{PyTorch} which uses automatic differentiation to compute the gradients~\cite{paszke2019pytorch}. 
The pseudocode for the training process is given in Algorithm \ref{alg:adam}. 
Further details on the pseudocode can be found in~\cite{kingma2015adam, goodfellow2016deep}.
For all experiments we used a learning rate scheduler that decays the learning rate exponentially after each epoch. 
The parameters for all models were initialized randomly from the uniform distribution over the interval $[-1,1]$. 
All feature scaling is performed using the \texttt{MinMaxScaler} class from the \texttt{scikit-learn} Python library~\cite{pedregosa2011scikit} with the exception of the handwritten digits dataset which was normalized by dividing the pixel intensities by 16. 
We did not find that regularization was necessary for our experiments but note that regularization of the displacement parameters, such as an $L_1$ or $L_2$ penalty, could be used to further improve energy efficiency, as we show in Appendix \ref{disp_reg}, as this is the only non-passive optical operation in the circuit. 
\begin{table}[ht]
\centering
\begin{tabular}{|c|c|}
\hline
Hyperparameter & Value \\
\hline
Learning rate ($\gamma$) & $10^{-2}$ \\
\hline
Learning rate decay ($\chi$) & $0.999$ \\
\hline
 $1$st moment estimates decay rate ($\beta_1$) & $0.9$ \\
\hline
$2$nd moment estimates decay rate ($\beta_2$) & $0.999$ \\
\hline
Number of epochs & $1000$ \\
\hline
Batch size & $32$ \\
\hline
\end{tabular}
\caption{Hyperparameters used when training the optical circuit for regression tasks. }
\label{tab:hyperparam_table_reg}
\end{table}

\begin{figure*}[htbp]
    \centering
    \begin{subfigure}[b]{0.32\textwidth}
        \centering
        \includegraphics[width=\textwidth]{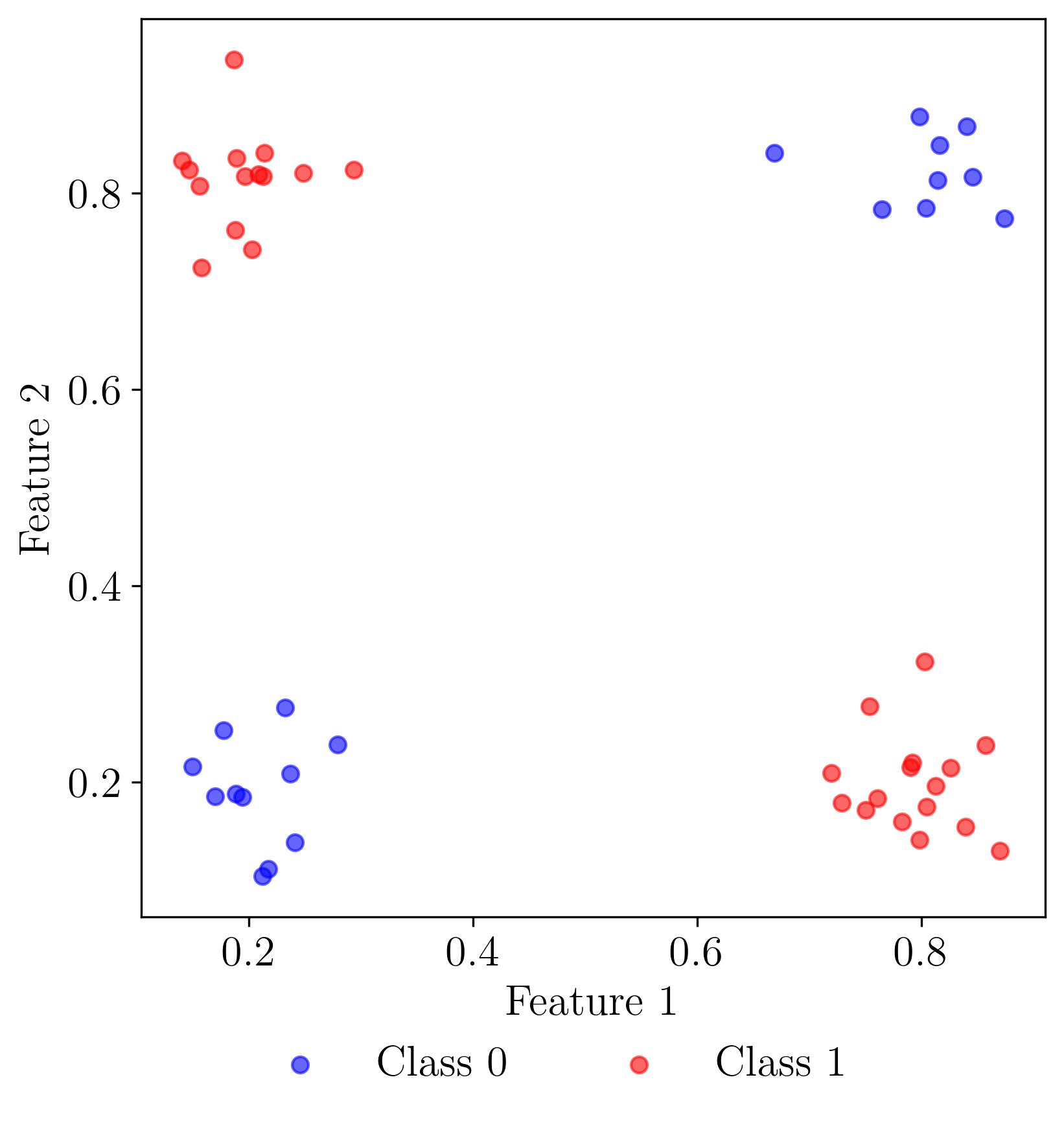}
        \caption{XOR classification problem.}
        \label{fig:xor_data}
    \end{subfigure}
    \begin{subfigure}[b]{0.32\textwidth}
        \centering
        \includegraphics[width=\textwidth]{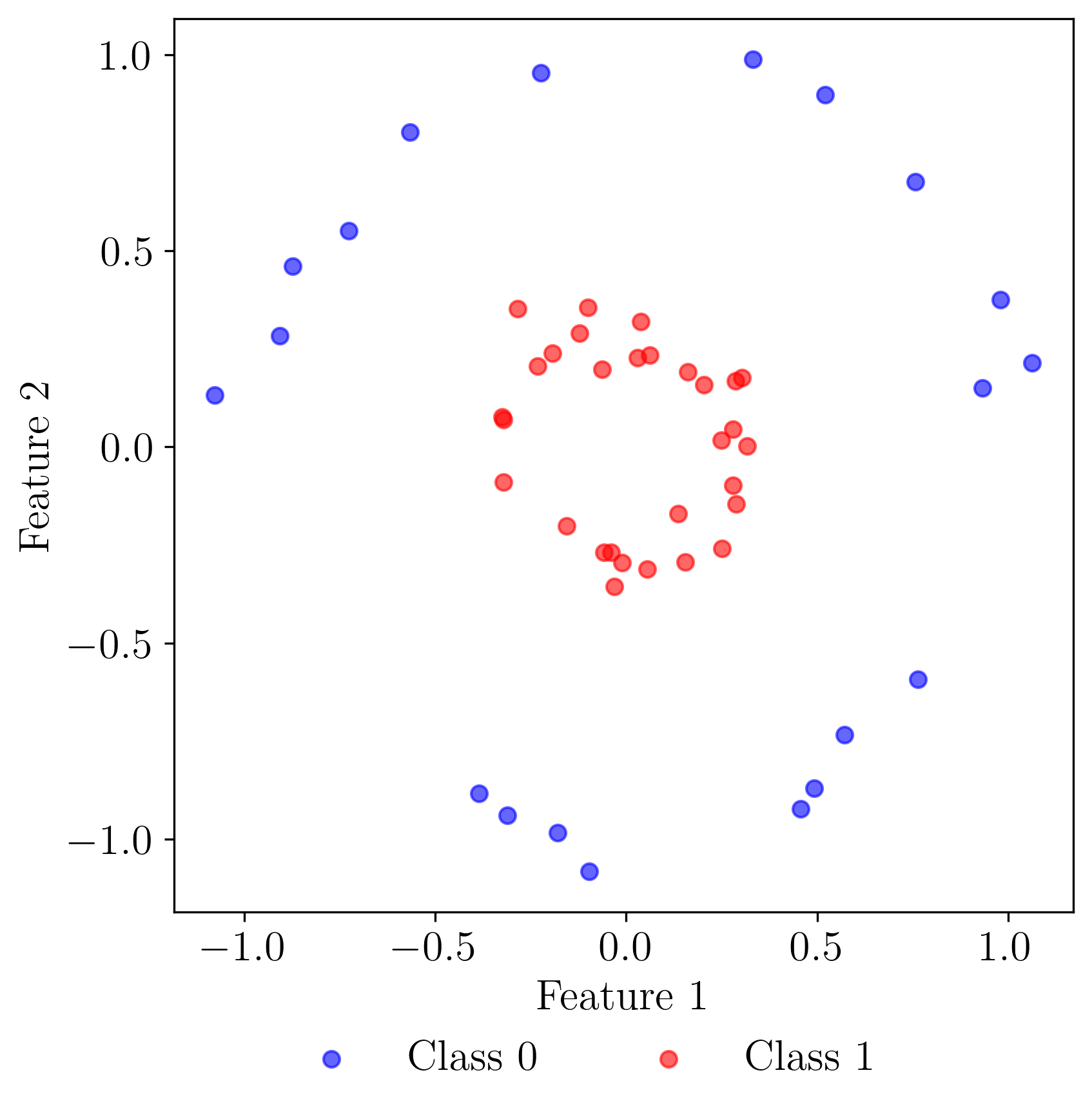}
        \caption{Circles classification problem.}
        \label{fig:circ_data}
    \end{subfigure}
    \begin{subfigure}[b]{0.32\textwidth}
        \centering
        \includegraphics[width=\textwidth]{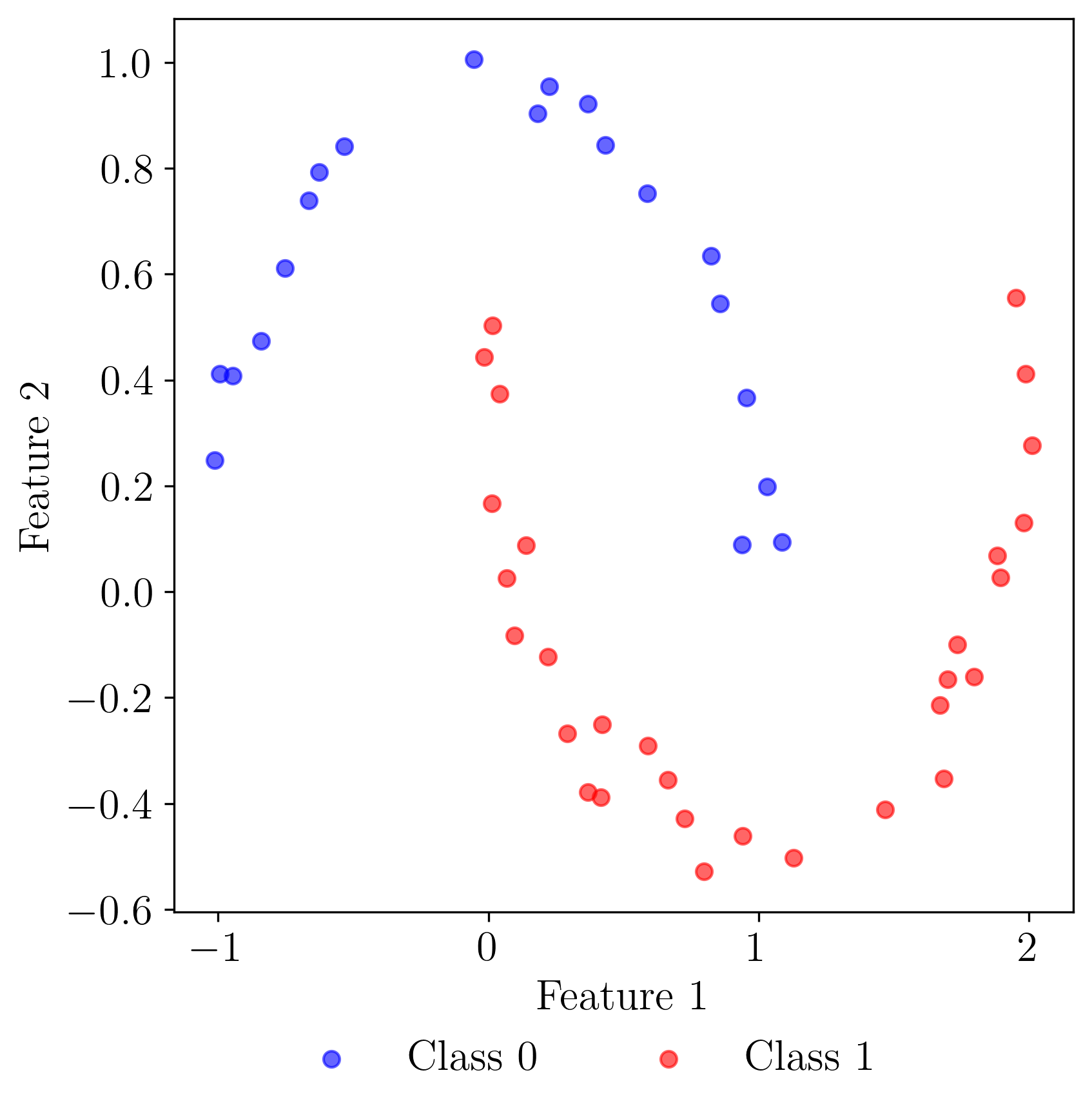}
        \caption{Moons classification problem.}
        \label{fig:mooons_data}
    \end{subfigure}
    \caption{Training data used for binary classification problems. We set the standard deviation of the Gaussian noise to $\sigma=0.05$. }
    \label{fig:2D-classification_data}
\end{figure*}

\subsection{Nonlinear regression}

The first supervised learning problem we investigated was that of nonlinear regression. In regression problems,  e.g.\ curve fitting, one attempts to learn a function which maps inputs to continuous outputs~\cite{james2023statistical}. Our circuit architecture used a three mode circuit with three layers.
The input scalar $x$ was repeatedly encoded in the coherent state amplitude of the three modes and in the phase angle of the phase shift operations on each mode in each subsequent layer.  
We fit four nonlinear functions: $\tanh(2x)$, $\sin(\pi x)$, $x^2$ and $x^3$.
We took the expectation value of the $Q$ quadrature of the first mode as our output. 
When generating the training data we added a Gaussian noise term to the ground truth labels with mean zero and standard deviation $\sigma=0.05$. 
We generated 100 samples and used 50 for training and 50 for testing.
The hyperparameters used for the regression training are shown in Table \ref{tab:hyperparam_table_reg}. For our loss function we used the mean squared error function as used in previous works~\cite{killoran2019continuous, bangar2023experimentally}.

The results of curve fitting on the noisy functions are shown in Fig.\ref{fig:lossless_regression}.
We can see that the predictions of the model on the test data points closely match the true noiseless function plotted in green for all functions.
This indicates that our nonlinear input encoding enables the network to learn a variety of nonlinear functions, even in the presence of noise, without the use of regularization techniques similar to previous CV quantum neural network architectures. 

We investigated the effect of photon loss in the circuit.
We defined the photon transmissivity $\eta\in[0,1]$ as $n_\text{out}=\eta \,n_\text{in}$ and hence the loss coefficient as $1-\eta$. In the particular case where all quantum light states are coherent states, the means of the quadratures transform classically via attenuation~\cite{cerf2007quantum}
\begin{align}
     q  &\mapsto \sqrt{\eta} \ q  \\
      p  &\mapsto \sqrt{\eta} \ p .
\end{align}
We apply these loss channels at the end of each Gaussian layer.
The results for fitting the function $x^3$ with a lossy circuit are shown in \fig{reg_lossy}. 
We can see that even with losses as high as $\eta=0.5$ the resulting fit was essentially unchanged. 
However, from the heatmap of displacement magnitudes on the right hand side we can see that larger displacements on the first mode were needed for most modes in the three layers to compensate for the higher losses.  

\begin{figure*}[htbp]
    \centering
    \begin{subfigure}[t]{0.9\textwidth}
    \begin{minipage}[b]{0.49\linewidth}
    \centering
    \includegraphics[width=\linewidth]{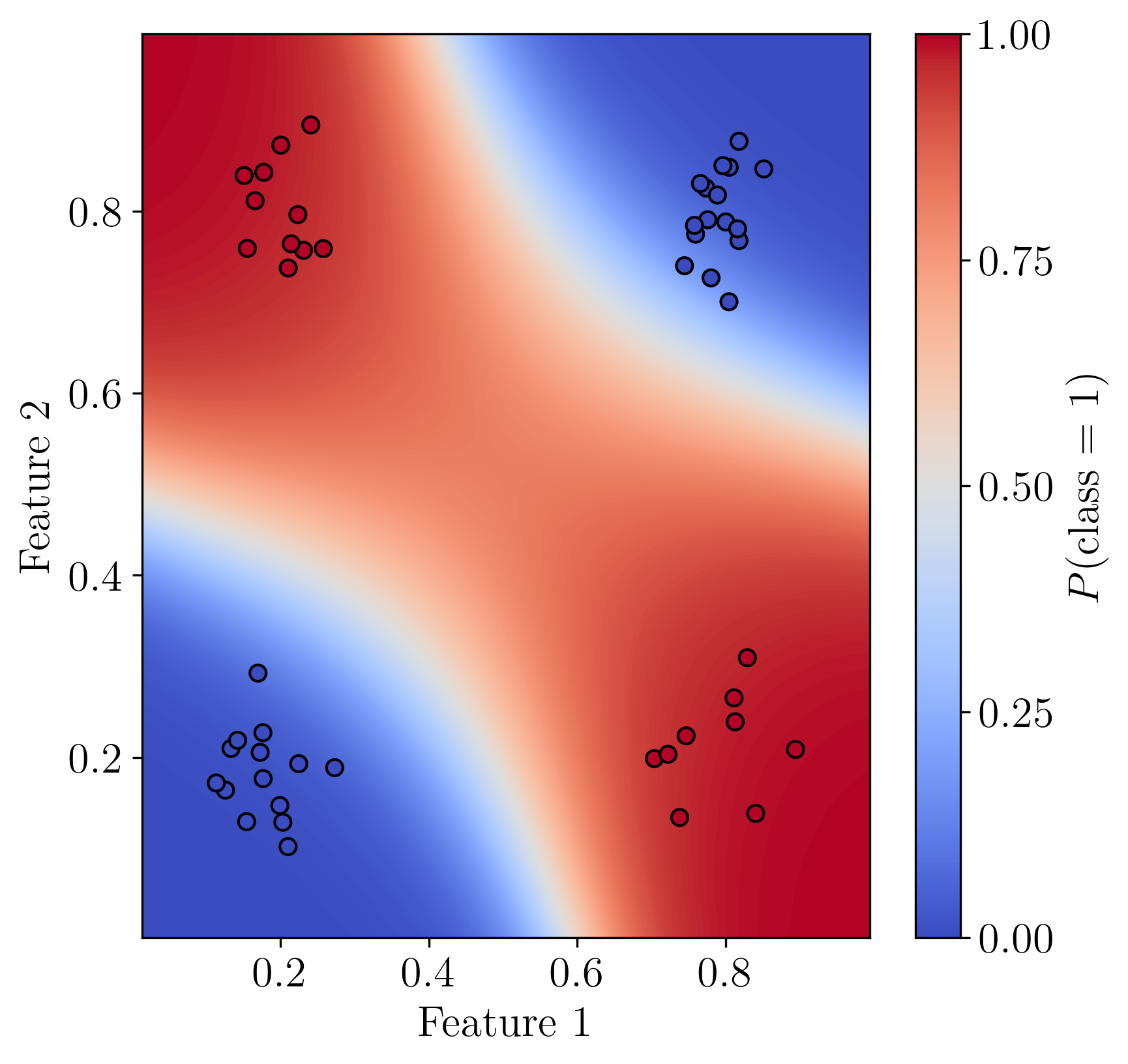}
    \end{minipage}\hfill
    \begin{minipage}[b]{0.49\linewidth}
      \centering
    \includegraphics[width=\linewidth]{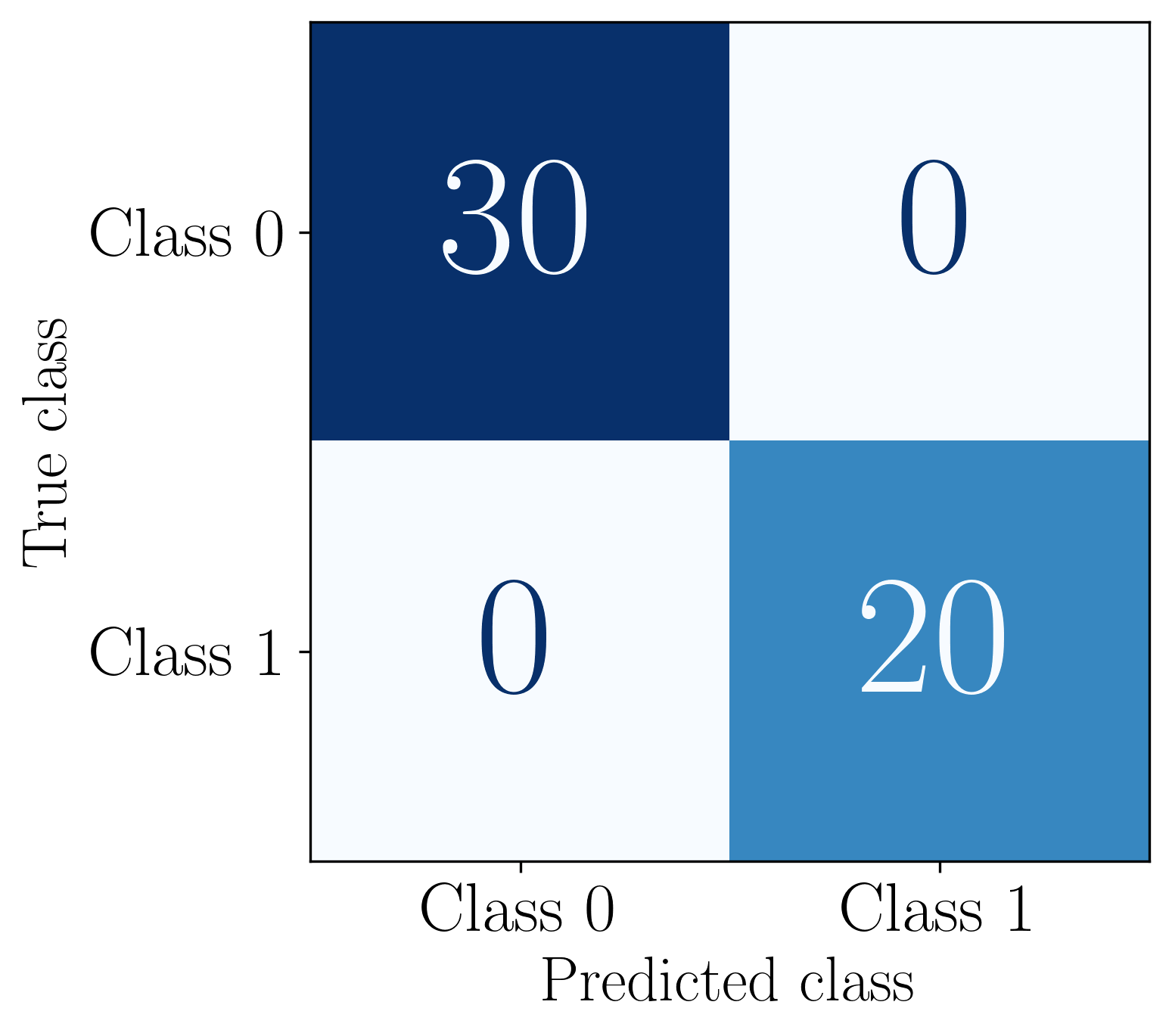}
    \end{minipage}
    \caption{Test set results for XOR classification problem.}
    \label{fig:xor_bin}
    \end{subfigure}
    \hfill
    \begin{subfigure}[t]{0.9\textwidth}
    \begin{minipage}[b]{0.49\linewidth}
    \centering
    \includegraphics[width=\linewidth]{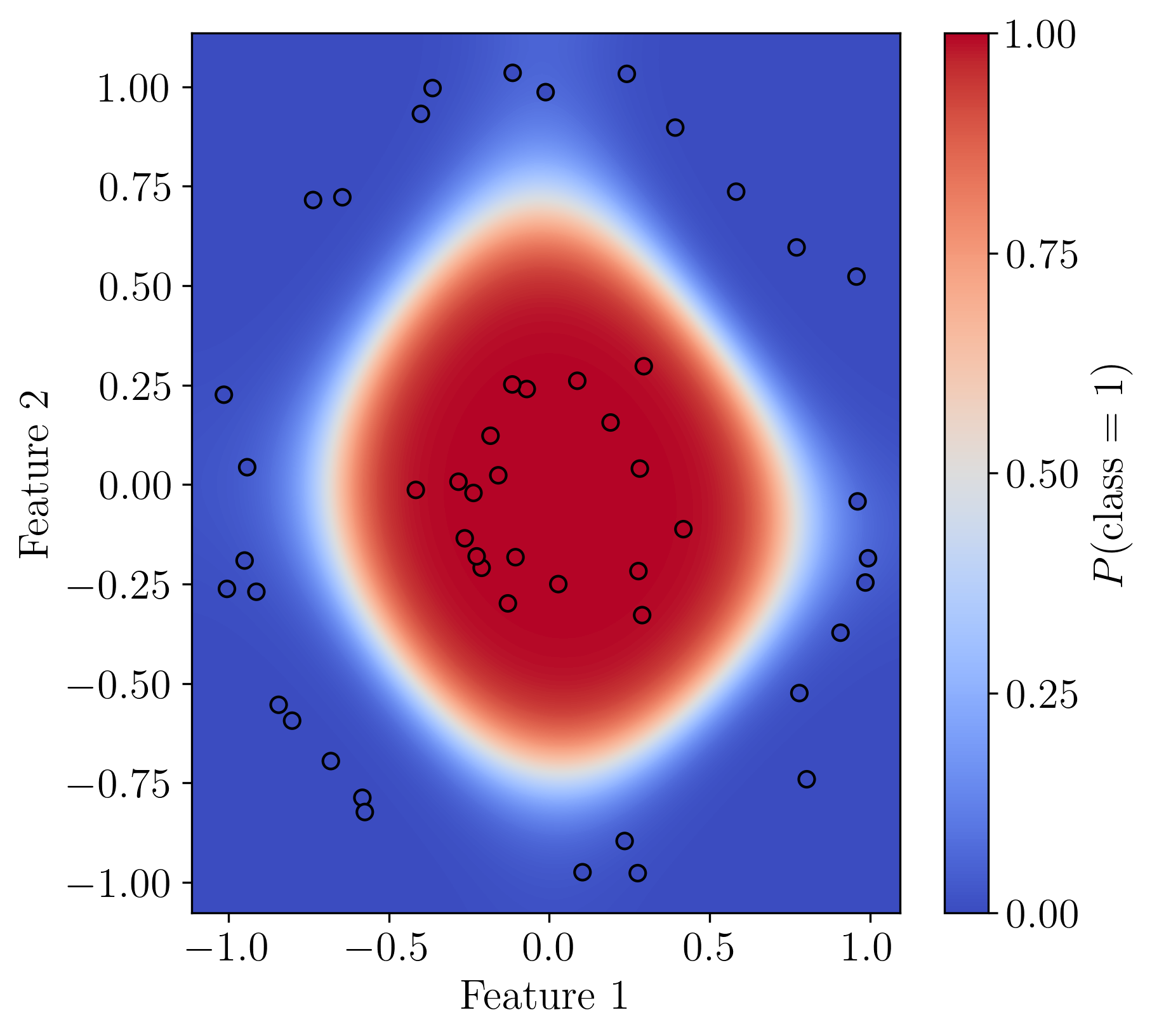}
    \end{minipage}\hfill
    \begin{minipage}[b]{0.49\linewidth}
      \centering
    \includegraphics[width=\linewidth]{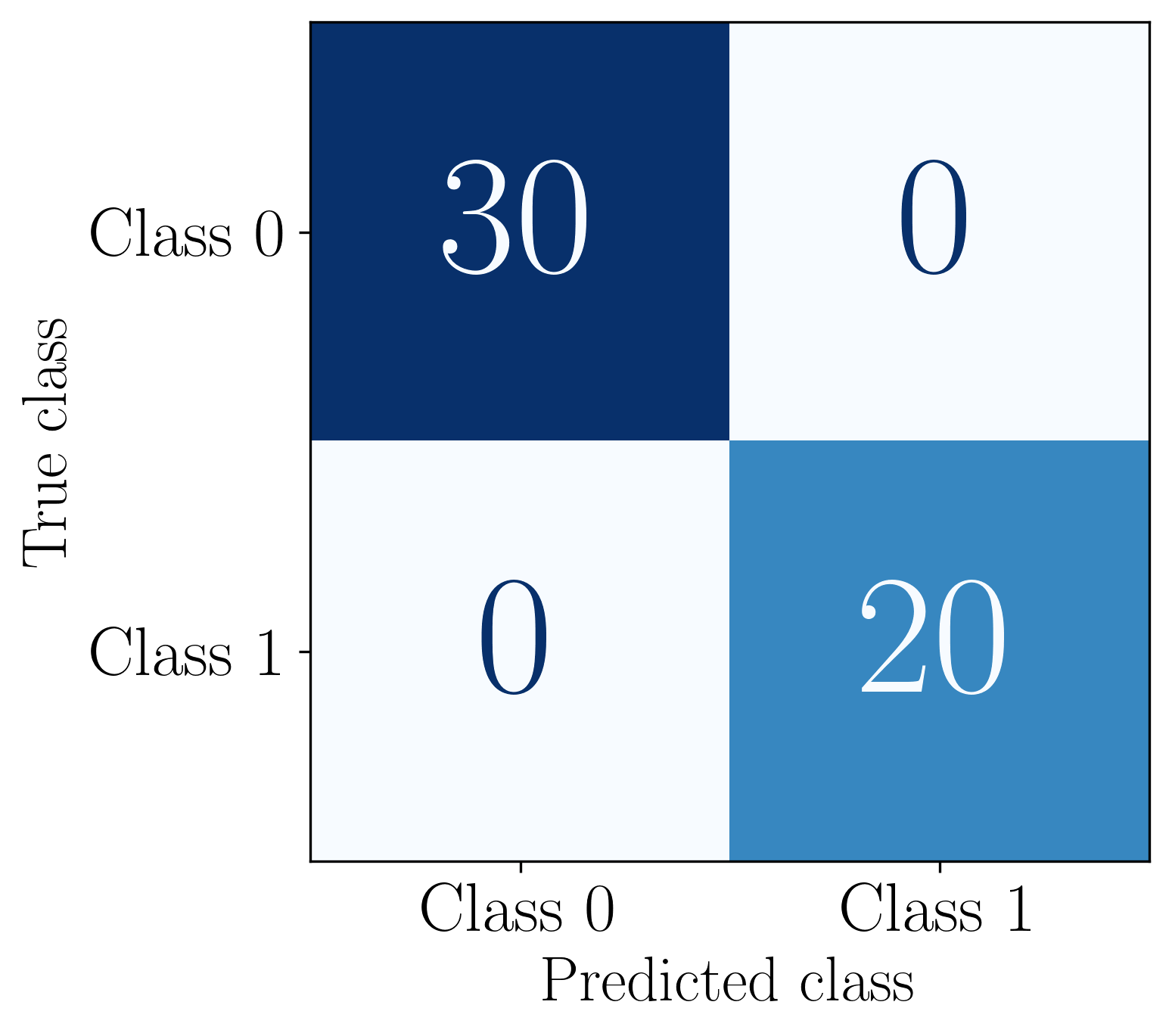}
    \end{minipage}
    \caption{Test set results for circles classification problem.}
    \label{fig:circ_bin}
    \end{subfigure}
    \hfill
    \begin{subfigure}[t]{0.9\textwidth}
    \begin{minipage}[b]{0.49\linewidth}
    \centering
    \includegraphics[width=\linewidth]{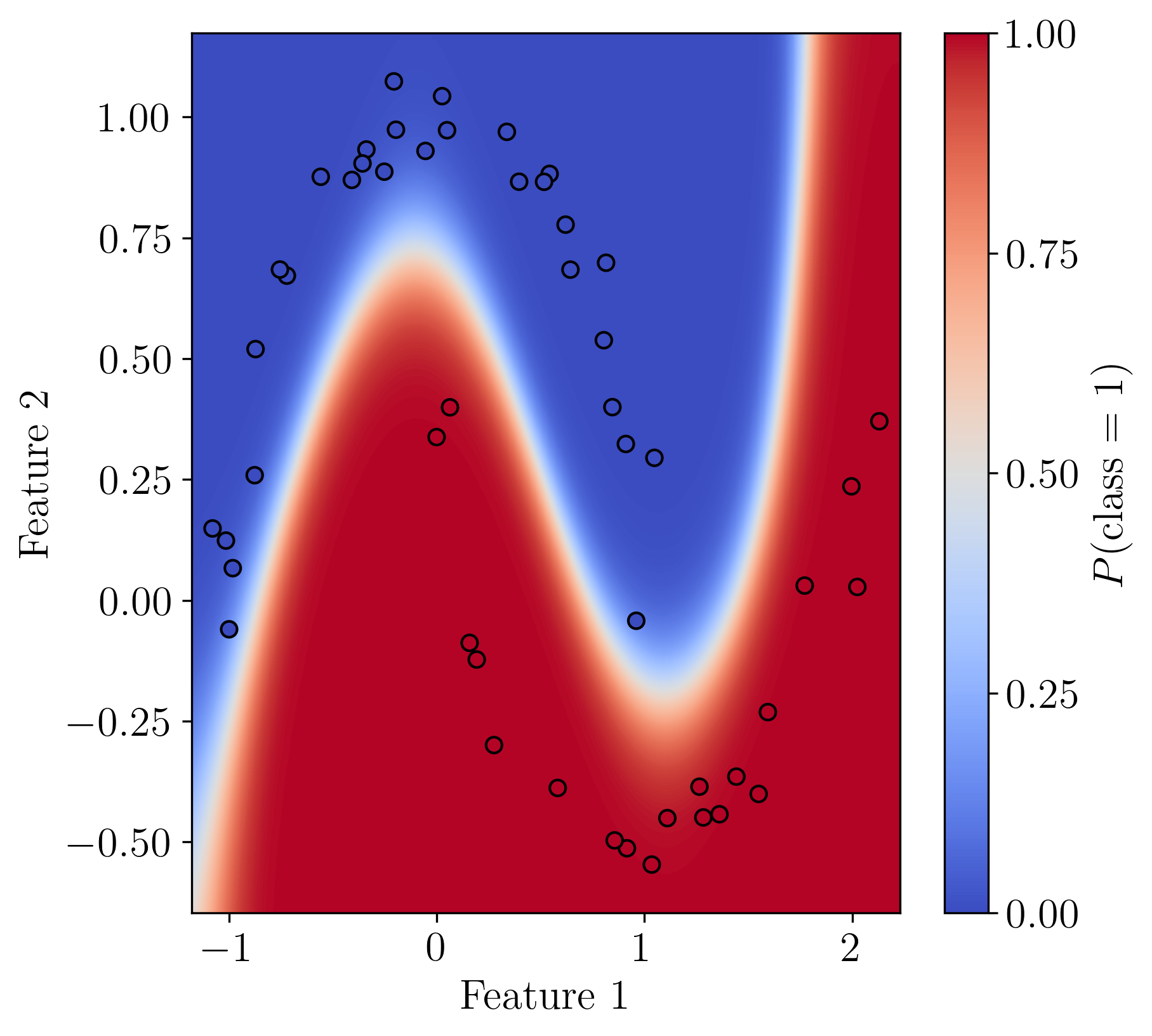}
    \end{minipage}\hfill
    \begin{minipage}[b]{0.49\linewidth}
      \centering
    \includegraphics[width=\linewidth]{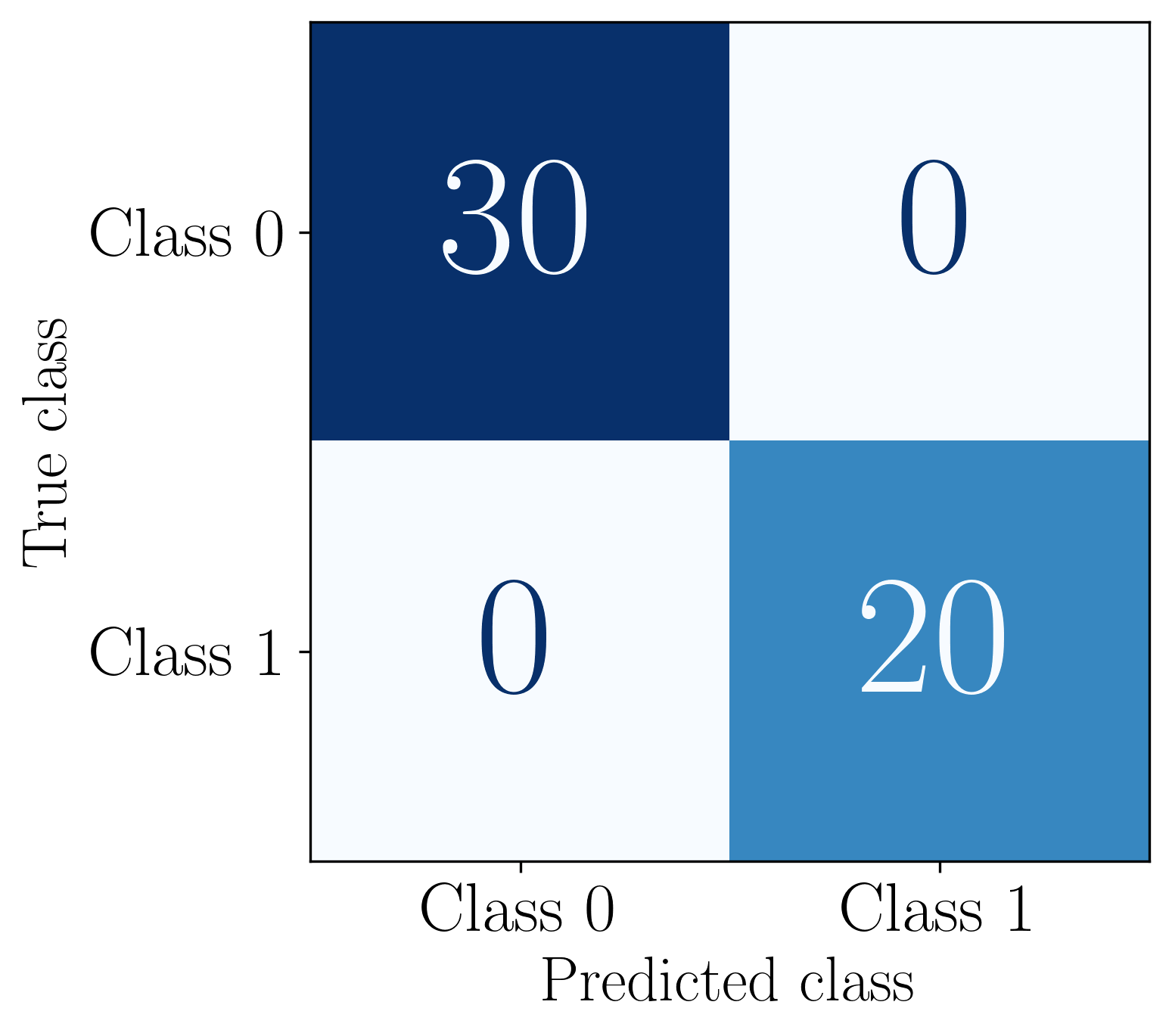}
    \end{minipage}
    \caption{Test set results for moons classification problem.}
    \label{fig:moon_bin}
    \end{subfigure}
    
    \caption{Classification results with ONN on the test set. The red data points correspond to class 1 and the blue correspond to class 0. The heatmap on the left shows the probability predicted by the network that a point in a particular region belongs to class 1 along with points from the test set. The confusion matrix for the test set is shown on the right.}
    \label{fig:2D-classification}
\end{figure*}

\begin{figure*}[htbp] 
  \centering
  \begin{subfigure}[t]{0.9\textwidth}
    \centering
    \begin{minipage}[b]{0.49\linewidth}
      \centering
      \includegraphics[width=\linewidth]{img/class/circ_noise_0.05_loss_0.0.png}
    \end{minipage}\hfill
    \begin{minipage}[b]{0.4\linewidth}
      \centering
      \includegraphics[width=\linewidth]{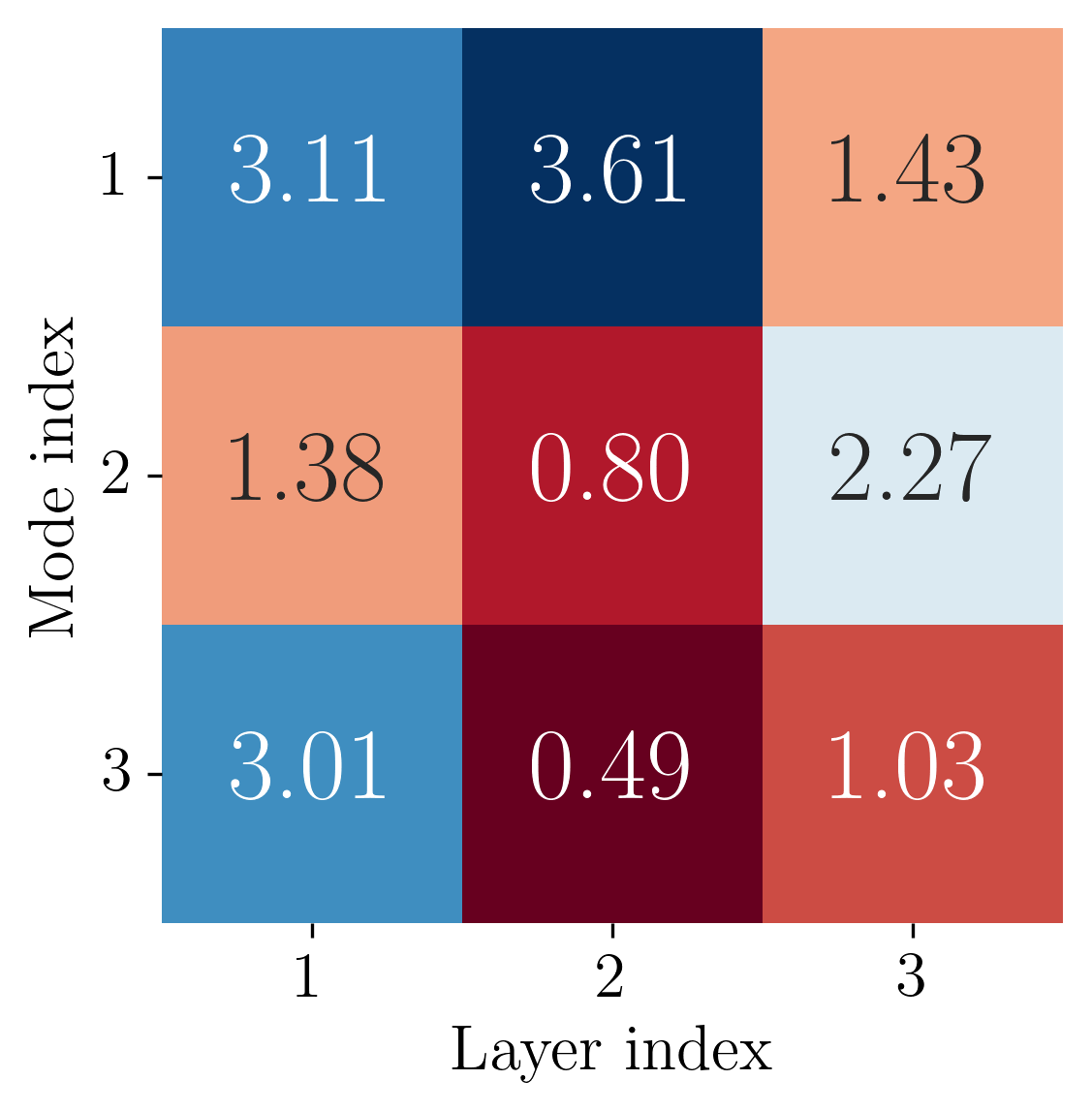}
    \end{minipage}
    \caption{Results with $\eta=0$.}
    \label{fig:class_lossy_sub1}
  \end{subfigure}\hfill
  \begin{subfigure}[t]{0.9\textwidth}
    \centering
    \begin{minipage}[b]{0.49\linewidth}
      \centering
      \includegraphics[width=\linewidth]{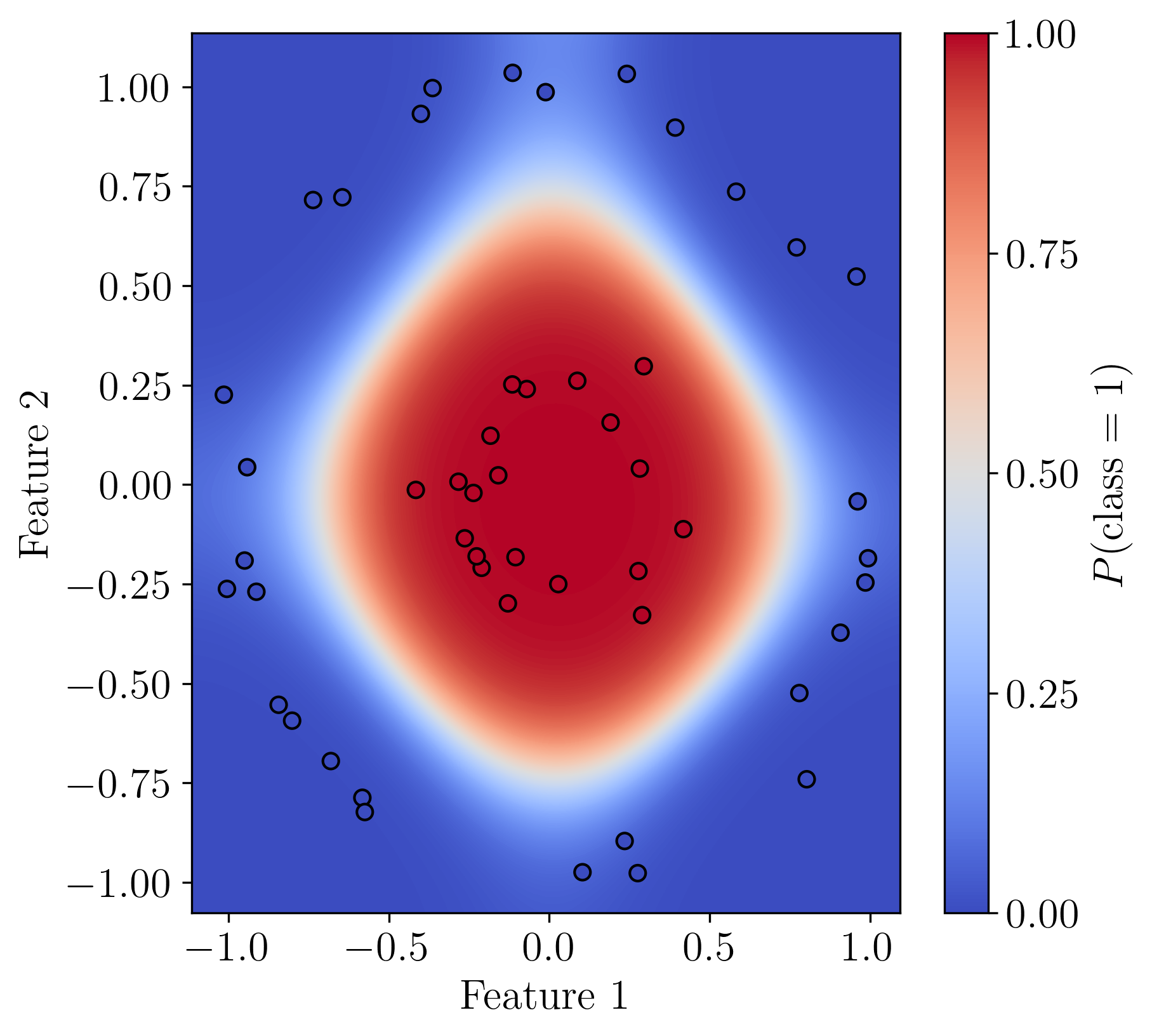}
    \end{minipage}\hfill
    \begin{minipage}[b]{0.4\linewidth}
      \centering
      \includegraphics[width=\linewidth]{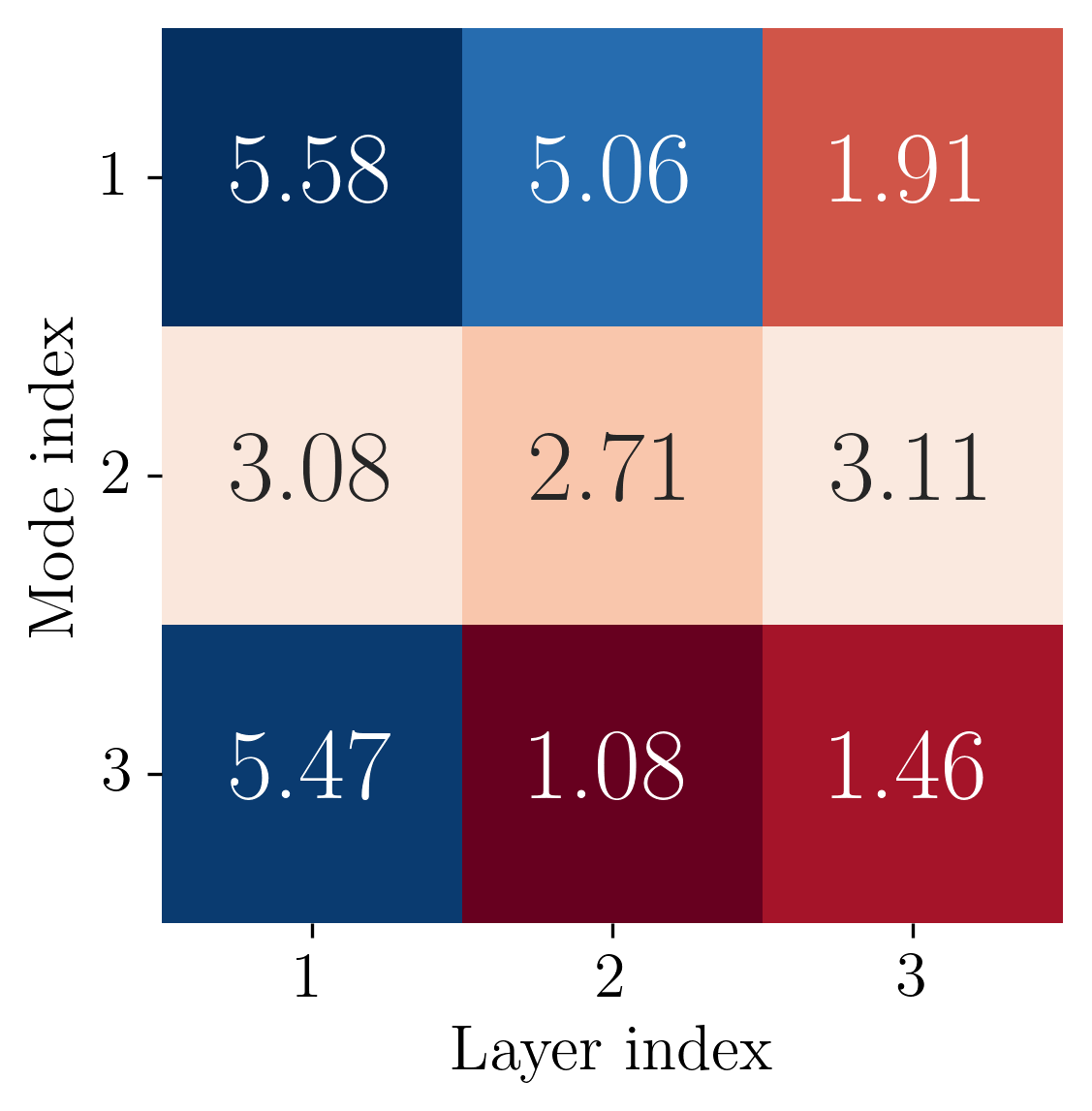}
    \end{minipage}
    \caption{Results with $\eta=0.5$.}
    \label{fig:class_lossy_sub2}
  \end{subfigure}
  \caption{Effect of photon loss after each layer of the optical circuit when learning to classify points on concentric circles. The heatmap on the right shows the magnitude $|\alpha|$ of the displacements applied to each mode at each layer of the optical circuit.}
  \label{fig:class_lossy}
\end{figure*}

\subsection{Classification}


We now examine how our model performs on classification which is the primary task used for benchmarking PNNs. For this class of supervised learning problems the model learns a function that maps inputs to discrete outputs, i.e., classes~\cite{james2023statistical}. 
For a classification problem with $C$ classes we treated the expectation value of the $Q$ quadrature of the first $C$ modes as the output logits of the model in contrast to previous CV quantum neural network proposals which used the expectation value of the photon number number observable measured using non-Gaussian photon number detection~\cite{killoran2019continuous, bangar2023experimentally}.
We used the cross entropy loss function, as done in Refs.~\cite{pai2023experimentally, wanjura2024fully, yildirim2024nonlinear, Bandyopadhyay2024, austin2025hybrid}, in particular we used the \texttt{PyTorch} implementation which internally normalizes the logits into probabilities by use of the softmax function~\cite{bishop2023deep}. 
Note that the use of the softmax function is needed only during model training and not during inference.

\subsubsection{Binary Classification}

\paragraph{Binary problems}

For binary classification we considered three problems. The first is a version of the classic XOR problem. 
The XOR problem has both theoretical and historical significance in the deep learning literature~\cite{goodfellow2016deep}. 
The problem consists of learning the exclusive OR (XOR) function which was proven to not be learnable by linear models such as single layer perceptrons. 
To generate data for our XOR classification problem we defined four cluster centers at Cartesian coordinates $(0.2,0.2)$ and $(0.8,0.8)$ for class 0 and $(0.8,0.2)$ and $(0.2,0.8)$ for class 1.
We then generated data points for each cluster by sampling them from a two-dimensional Gaussian with independent components with means equal to the cluster center and standard deviation $\sigma=0.05$.
The other two problems that were considered consisted of learning a decision boundary for classes of points that lie on two concentric circles and that lie on two interleaving half circles (`moons') on the Cartesian plane as done in~\cite{pai2023experimentally}. 
Importantly, no linear model—that is, a model whose decision boundary in feature space is a hyperplane—can partition the feature space of any of these three problems such that all data points belonging to a given class lie on a single side of the boundary.

We generated data points for the last two problems using \texttt{scikit-learn} with the same Gaussian noise as in the XOR case. 
We generated 100 data points and used a 50/50 split of the data into training and test sets for all problems. 
The hyperparameters are given in \tab{hyperparam_table_class}.
We used a three-mode circuit with three layers for all binary classification problems and encode the inputs into the first two modes with the remaining mode being vacuum. 
The output logits were taken to be the expectation of the $Q$ quadrature of the first two modes of the circuit.

\begin{table}[ht]
\centering
\begin{tabular}{|c|c|}
\hline
Hyperparameter & Value \\
\hline
Learning rate ($\gamma$) & $10^{-2}$ \\
\hline
Learning rate decay ($\chi$) & $0.999$ \\
\hline
 $1$st moment estimates decay rate ($\beta_1$) & $0.9$ \\
\hline
$2$nd moment estimates decay rate ($\beta_2$) & $0.999$ \\
\hline
Number of epochs & $1000$ \\
\hline
Batch size & $32$ \\
\hline
\end{tabular}
\caption{Hyperparameters used when training the optical circuit for binary classification tasks. }
\label{tab:hyperparam_table_class}
\end{table}
The training data for each of the problems are shown in Fig.\ref{fig:2D-classification_data}.

The results of evaluating the trained models on the test sets are visualized using the class probability heatmaps and the confusion matrices in Fig.\ref{fig:2D-classification}. For all classification problems we achieved a test accuracy of 100\% as can be seen from the confusion matrices on the right. 

\paragraph{Lossy case}

Fig.\ref{fig:class_lossy} shows the performance of the model when trained with no photon loss and with photon loss where $\eta=0.5$ for the concentric circles problem.
We can see that, as in the regression case, the model could still learn an appropriate nonlinear decision boundary in the presence of high losses. 
From the accompanying heatmaps we observed that, similar to the regression case larger displacements were applied in the lossy case. 
However, in contrast to the regression example all displacements throughout the network were higher in the lossy case, not just for the modes and layers closer to the output homodyne measurements. 
\begin{figure*}[htbp]
    \centering
    \begin{subfigure}[b]{0.32\textwidth}
        \centering
        \includegraphics[width=\textwidth]{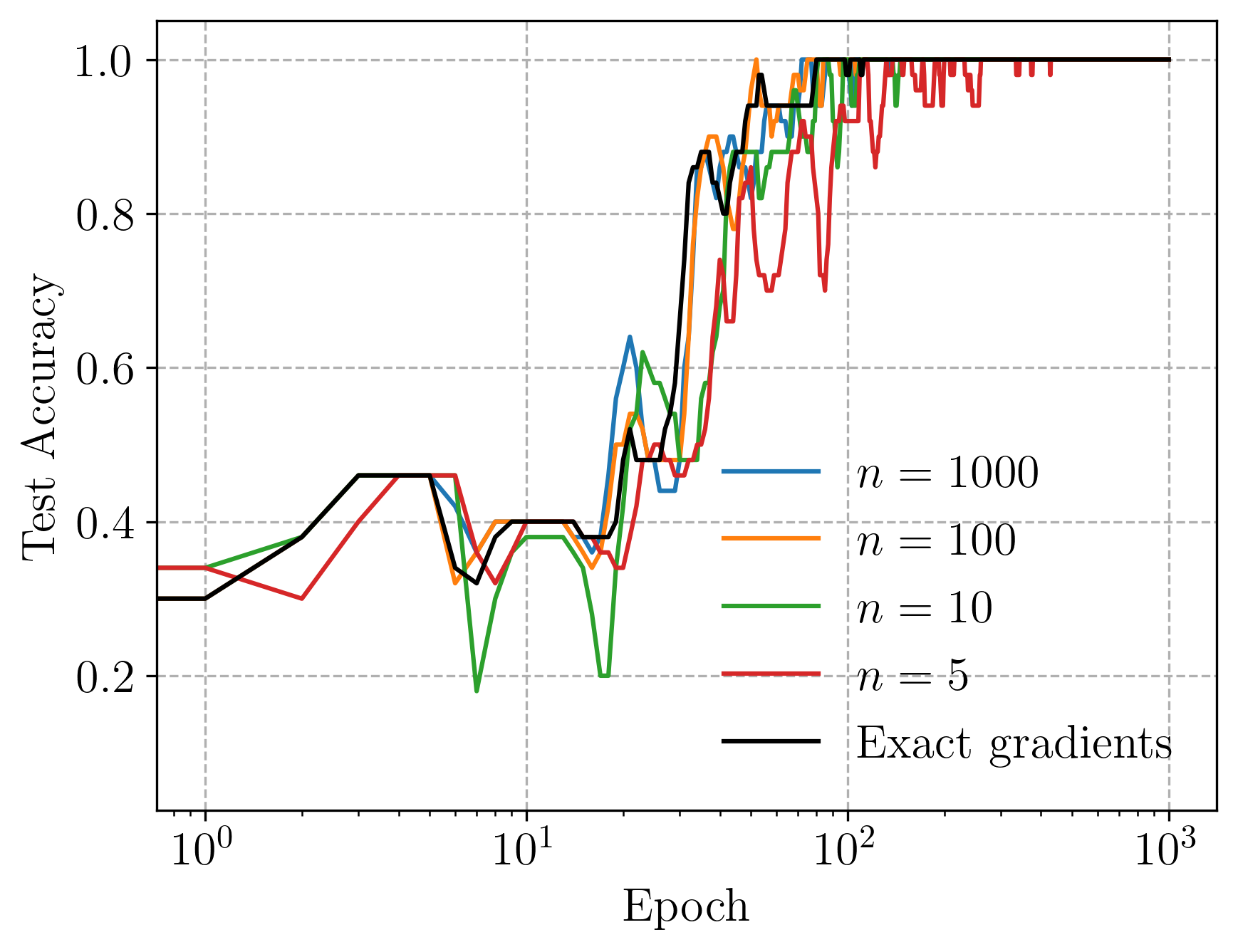}
        \caption{Learning curve for the XOR problem.}
        \label{fig:shot_noise_xor}
    \end{subfigure}
    \begin{subfigure}[b]{0.32\textwidth}
        \centering
        \includegraphics[width=\textwidth]{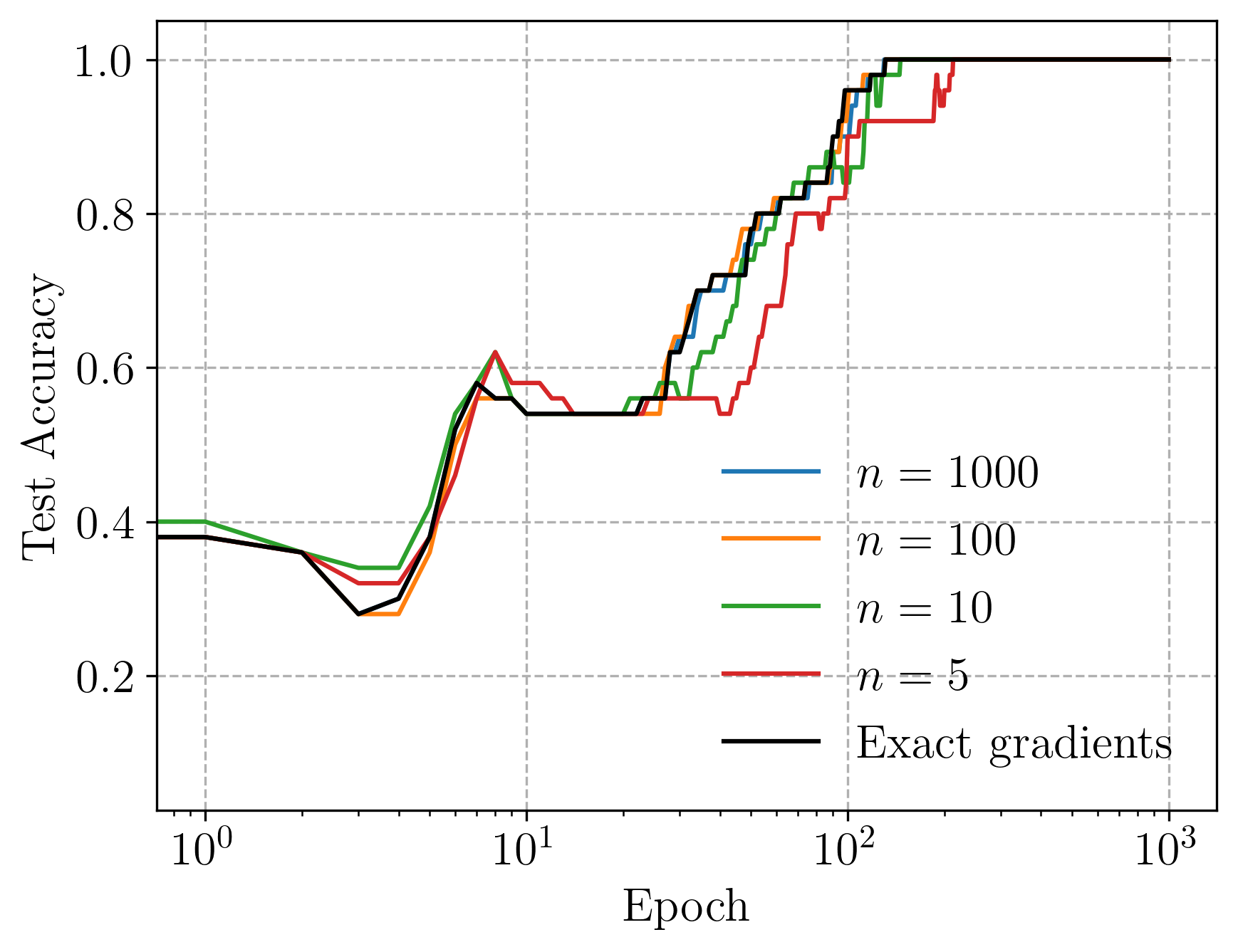}
        \caption{Learning curve for the concentric circles problem.}
        \label{fig:shot_noise_circ}
    \end{subfigure}
    \begin{subfigure}[b]{0.32\textwidth}
        \centering
        \includegraphics[width=\textwidth]{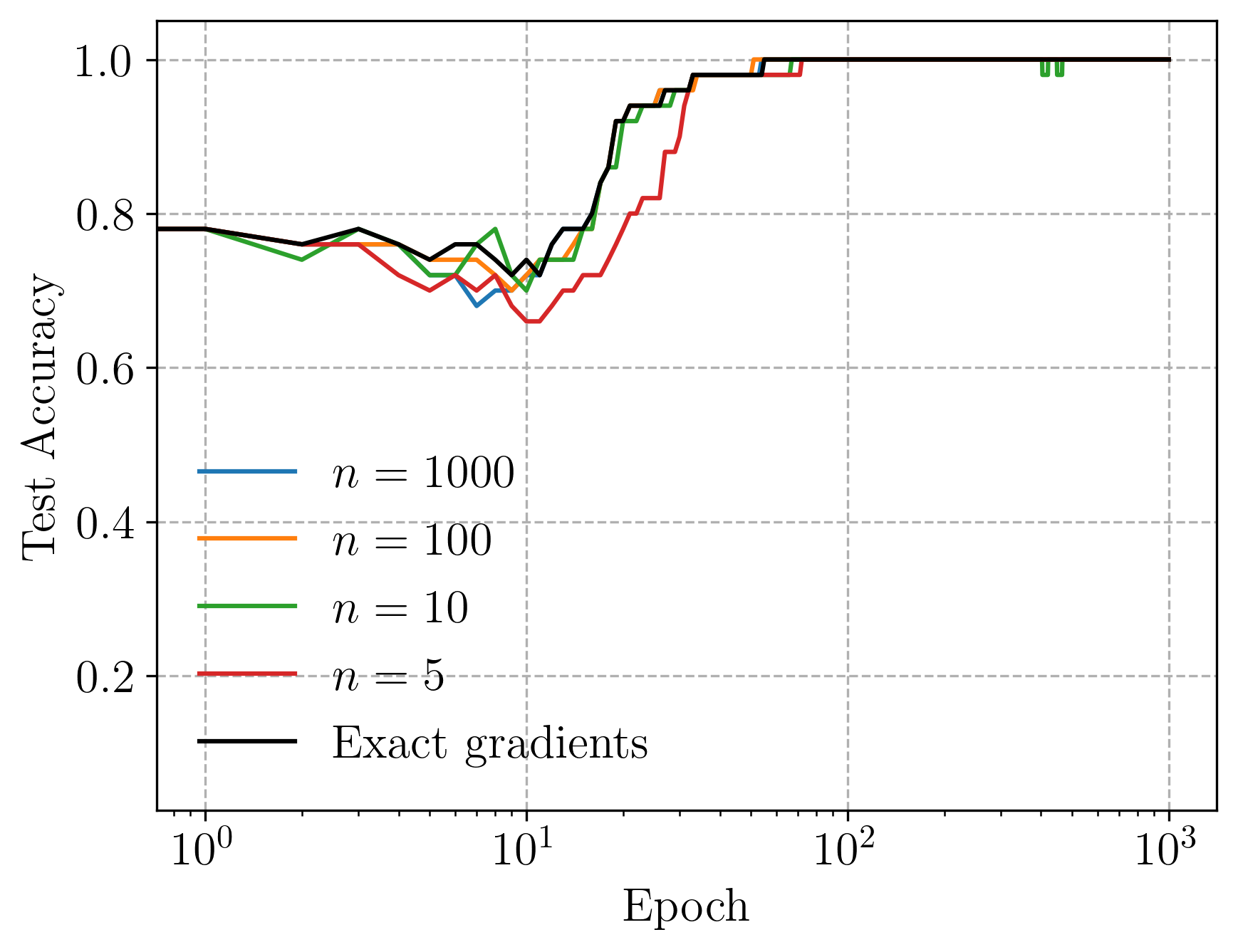}
        \caption{Learning curve for the moons problem.}
        \label{fig:shot_noise_moon}
    \end{subfigure}
    \caption{Plots of model accuracy for binary classification problems on the test set over the course of training when using various values of $n$ to estimate gradients.}
    \label{fig:shot_noise}
\end{figure*}

\paragraph{Hardware-based gradient determination}
We investigated the effect on learning from the variance in parameter-shift-rule gradient estimation.
As recalled above, the parameter shift rules require subtraction of expectation values which we would estimate in an actual experiment using the sample mean. 
In our case since the quadratures are normally distributed random variables their sample means are also normally distributed.   
As is well known~\cite{lemons2002introduction}, for two independent normally distributed random variables $X$ and $Y$ with means $\mu_X$ and $\mu_Y$  and variances $\sigma^2_X$ and $\sigma^2_Y$, the random variable $Z=X-Y$ is also normally distributed with mean $\mu_X-\mu_Y$ and variance $\sigma^2_X+\sigma^2_Y$.
Using this we can simulate the noise from using the parameter shift rules by using an estimate of the true gradients when performing stochastic gradient descent.
These estimates are computed by drawing $n$ samples from a Gaussian with mean equal to the true gradient, as computed by automatic differentiation, and variance $\sigma^2=1/n$ and then computing the sample mean.
The results for solving the three binary classification problems using various numbers of shots are shown in Fig.\ref{fig:shot_noise}.
We see that all models still reach $100\%$ test accuracy for various numbers of samples being used. 
However, we find that as less samples are used slightly more epochs are needed for the models to converge to the best test accuracy and that the stability of the convergence is lower. 
This is in agreement with findings for qubit based quantum circuits \cite{sweke2020stochastic}.
Notably, the plots indicate that just $100$ shots are often enough to reach stable convergence with comparable convergence time to the exact gradient case.

\subsubsection{Multilabel Classification}
\label{sec:multilabel}

In addition to the previous sections which demonstrate this ONN model can indeed solve non-trivial nonlinear learning problems we also investigate the performance of our network on more challenging and commonly used benchmarks of multilabel classification. 
We used three examples of multilabel classification: the Iris data set  which is a commonly used benchmark~\cite{anisetti2023learning, anisetti2024frequency, zhang2021optical, li2024training, ji2025photonic, xue2024fully}, the handwritten digits dataset considered in Refs.~\cite{wang2024training, wanjura2024fully} and the vowel recognition dataset which is widely used in benchmarking of physical neural networks \cite{shen2017deep, romera2018vowel, hughes2019wave, wright2022deep, momeni2023backpropagation, Bandyopadhyay2024, onodera2025arbitrary}. 
The first two datasets are loaded via \texttt{scikit-learn}.

\paragraph{Iris dataset}

\begin{figure}[htbp]{}
        \centering
        \includegraphics[width=.8\columnwidth]{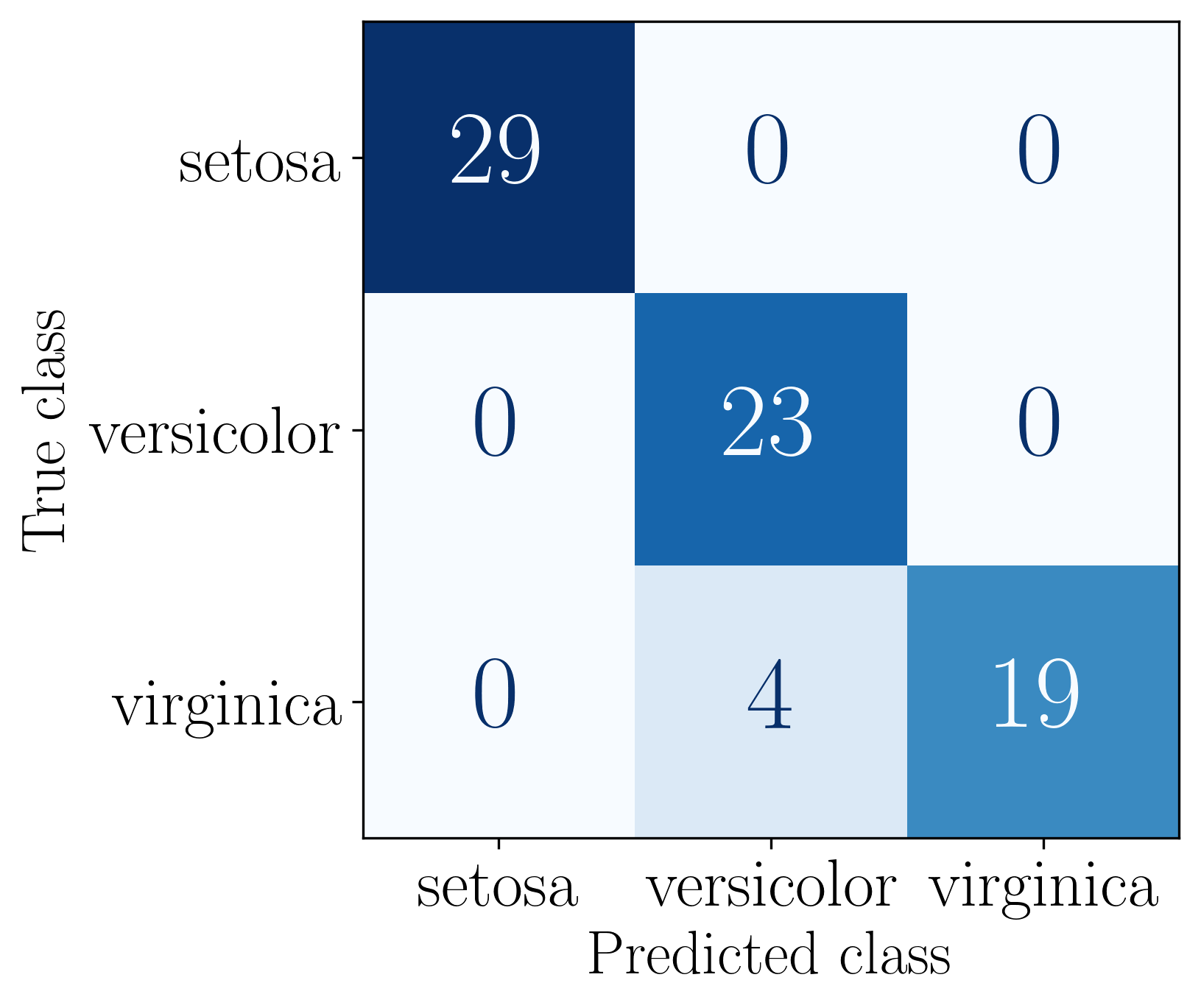}
        \caption{Confusion matrix for ONN classifier on test set of the Iris dataset.}
        \label{fig:cm_iris}
\end{figure}

\begin{figure*}[!htbp]{}
        \centering
        \includegraphics[width=.9\textwidth]{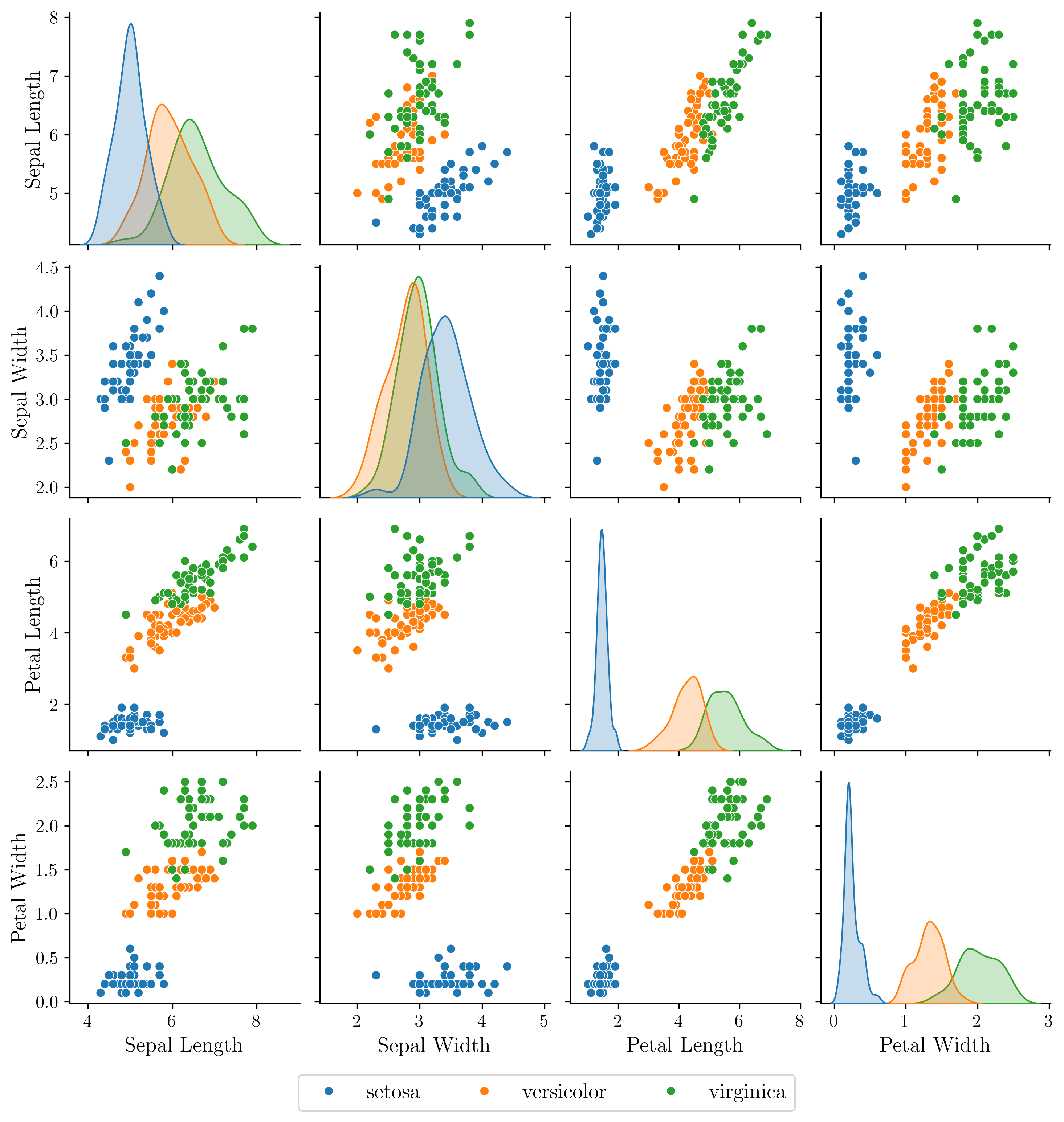}
        \caption{Pair plot of Iris dataset.}
        \label{fig:pair_plot_iris}
\end{figure*}

The Iris dataset consists of 150 samples from three different species (classes) of Iris flower (setosa, virginica, and versicolor) with each sample being represented as a four dimensional vector with elements corresponding to petal width, petal length, sepal width and sepal length in cm~\cite{fisher1936use}. 
We used 75 random samples for training and the rest for testing and normalized each input feature to be on the range $[0,1]$. 
We used a four mode circuit with three layers, leading to 42 trainable parameters, and the same hyperparameters used in the binary classification section. 
We encode the four features into the four modes as done in our previous experiments.
The expectation value of the $Q$ quadrature of the first three modes are taken to be our models output logits.
Fig.\ref{fig:cm_iris} shows the confusion matrix of the ONN evaluated on the test set. 
We can see that the test accuracy is $94.67\%$. 
The two species the model struggles slightly to differentiate are virginica and versicolor. 
This is in accordance with previous findings using this dataset in the literature~\cite{ji2025photonic} as the two species have significant overlap in feature space particularly with their sepal widths and sepal lengths as shown in the pair plot in Fig.\ref{fig:pair_plot_iris}. 

\begin{table}[ht]
\centering
\begin{tabular}{|c|c|}
\hline
Hyperparameter & Value \\
\hline
Learning rate ($\gamma$) & $10^{-2}$ \\
\hline
Learning rate decay ($\chi$) & $0.999$ \\
\hline
 $1$st moment estimates decay rate ($\beta_1$) & $0.9$ \\
\hline
$2$nd moment estimates decay rate ($\beta_2$) & $0.999$ \\
\hline
Number of epochs & $500$ \\
\hline
Batch size & $512$ \\
\hline
\end{tabular}
\caption{Hyperparameters used when training the optical circuit for the digits classification task. }
\label{tab:hyperparam_table_digits}
\end{table}

\begin{figure*}[htbp]{}
        \centering
        \includegraphics[width=.8\textwidth]{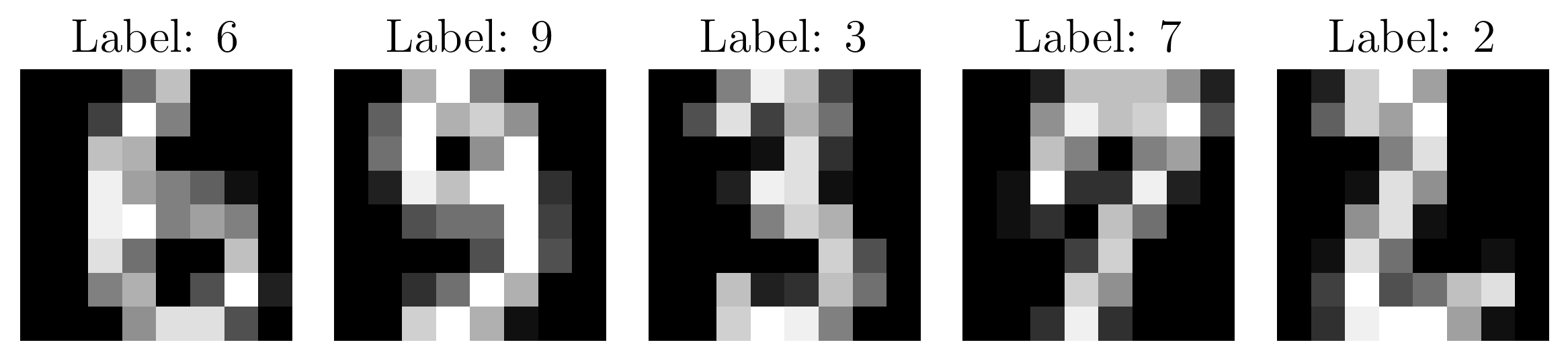}
        \caption{Examples of handwritten digits and their corresponding ground truth labels.}
        \label{fig:digits_examples}
\end{figure*}

\paragraph{Handwritten digits dataset}

\begin{figure}[htbp]
        \centering
        \includegraphics[width=.8\columnwidth]{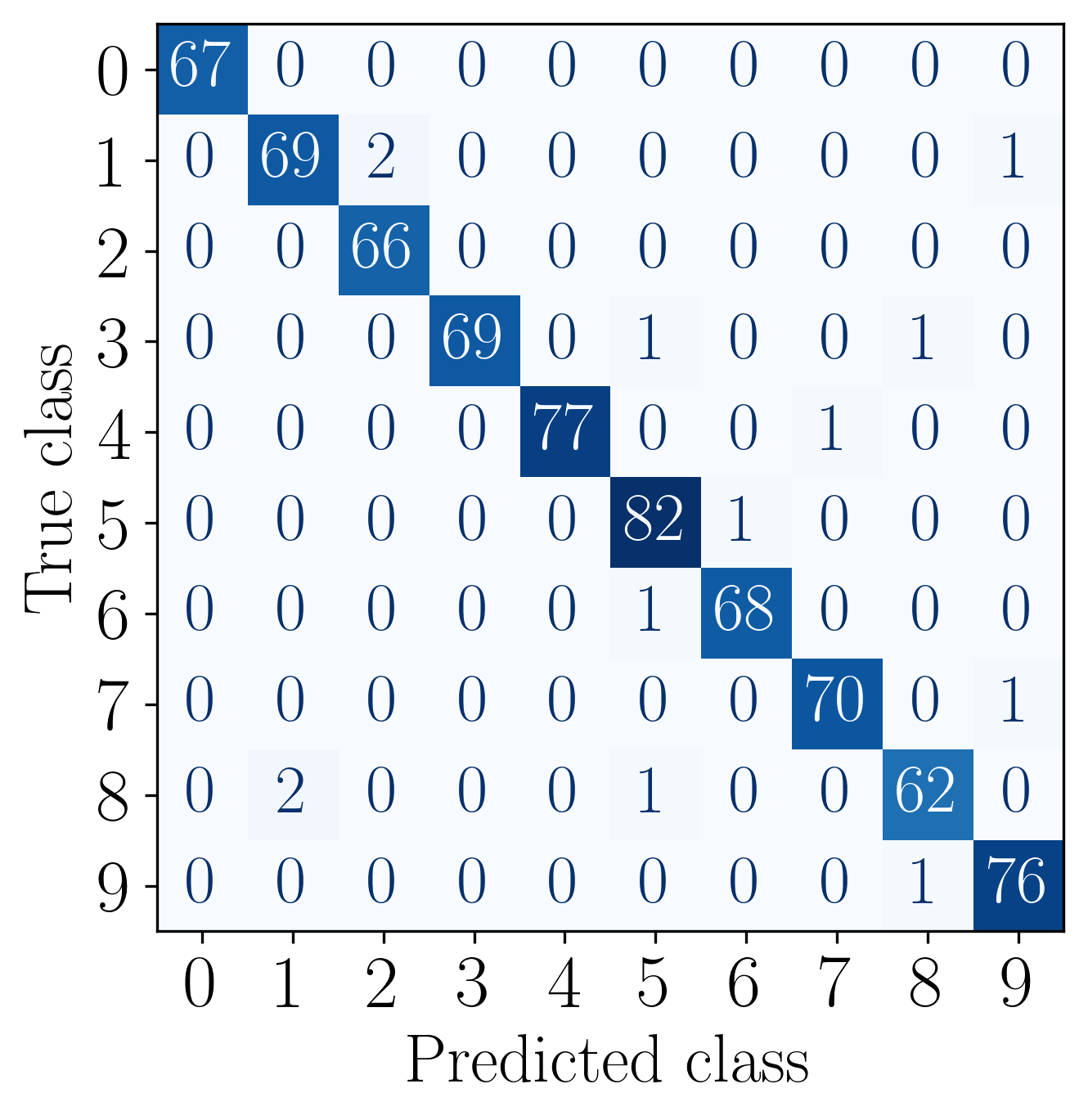}
        \caption{Confusion matrix for ONN classifier on test set of the handwritten digits dataset.}
        \label{fig:cm_digits}
\end{figure}

The handwritten digits dataset consists of 1797 gray scaled images of handwritten digits from 0 to 9 and was considered. 
For visualization purposes we show five example images in Fig.\ref{fig:digits_examples}.
We used 60\% of the data (1078 images) for training and 40\% (719 images) for testing. 
The pixel values were normalized to be between $[0,1]$ and the images were flattened into a $64$ dimensional vector. 
For this problem we used a 64 mode circuit with four layers, leading to $8576$ trainable parameters, and took the expectation value of the $Q$ quadrature of the first 10 modes as the output logits of the model. We encode the 64 normalized pixel intensities into the 64 modes of the optical circuit. 
The hyperparameters used to train the ONN for this task are shown in Table \ref{tab:hyperparam_table_digits}.
We achieved a test accuracy of 98.19\%  as can be seen by examining the confusion matrix of the trained model in Fig.\ref{fig:cm_digits}.

For comparison purposes, we trained a classical neural network with ReLU activation and one hidden layer of size $64$---so the number of parameters was comparable to the number in our model at $8970$---and obtained similar results with a test accuracy of $96.94\%$.

\begin{figure}[htbp]
        \centering
        \includegraphics[width=.8\columnwidth]{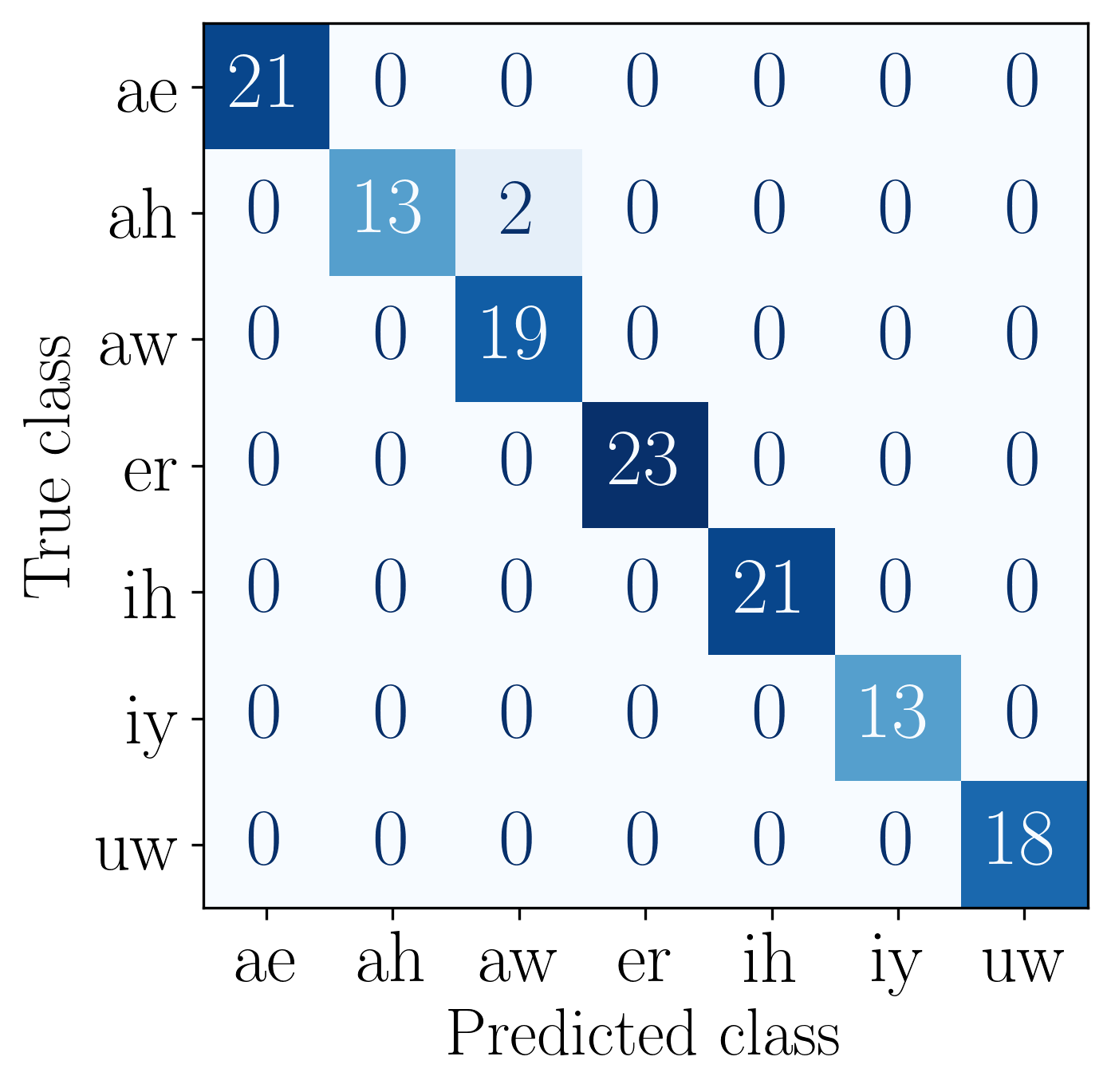}
        \caption{Confusion matrix for ONN classifier on test set of the vowel recognition dataset.}
        \label{fig:cm_vowel}
\end{figure}

\paragraph{Vowel recognition dataset} 
We consider the vowel recognition dataset from Refs.~\cite{romera2018vowel, wright2022deep, momeni2023backpropagation, onodera2025arbitrary},
which is a subset of the dataset used in Ref.~\cite{hillenbrand1995acoustic}, that consists of pronunciations of seven vowels by 37 female speakers where each vowel is characterized by 12 different formant frequencies. 
Specifically, each data point in the dataset is represented as a 12-dimensional vector containing the first three formants $F_1, F_2$ and $F_3$ sampled at the steady state, $20\%$, $50\%$, and $80\%$ of the vowels pronunciation duration respectively. 
Each speaker pronounces each of the seven vowels leading to a dataset with 259 data points.
We use a 12 mode circuit with three layers, leading to 270 trainable parameters, and the same hyperparameters as in the binary classification section with the learning rate set to $10^{-3}$. 
We use a random split of the data into 129 training samples and 130 testing samples.
We normalize the 12 features such that they are all between $[0,1]$.
From the confusion matrix in Fig. \ref{fig:cm_vowel} we can see that the ONN achieves a test accuracy of 98.46\%.

\begin{figure*}[htbp]
    \centering
    \begin{subfigure}[b]{0.38\textwidth}
        \centering
\includegraphics[width=\textwidth]{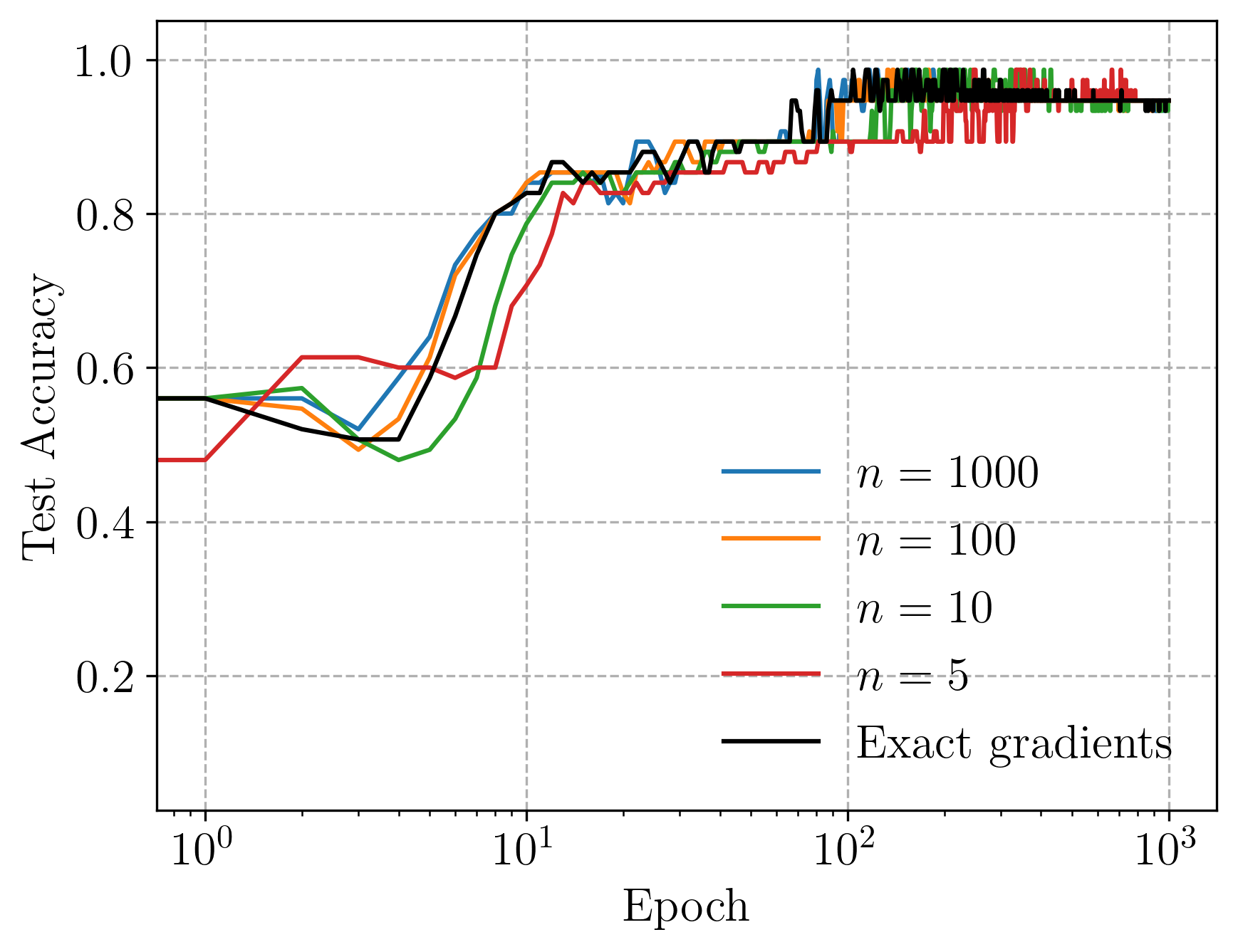}
        \caption{Learning curve for the Iris dataset.}
        \label{fig:shot_noise_iris}
    \end{subfigure}
    \begin{subfigure}[b]{0.38\textwidth}
        \centering
        \includegraphics[width=\textwidth]{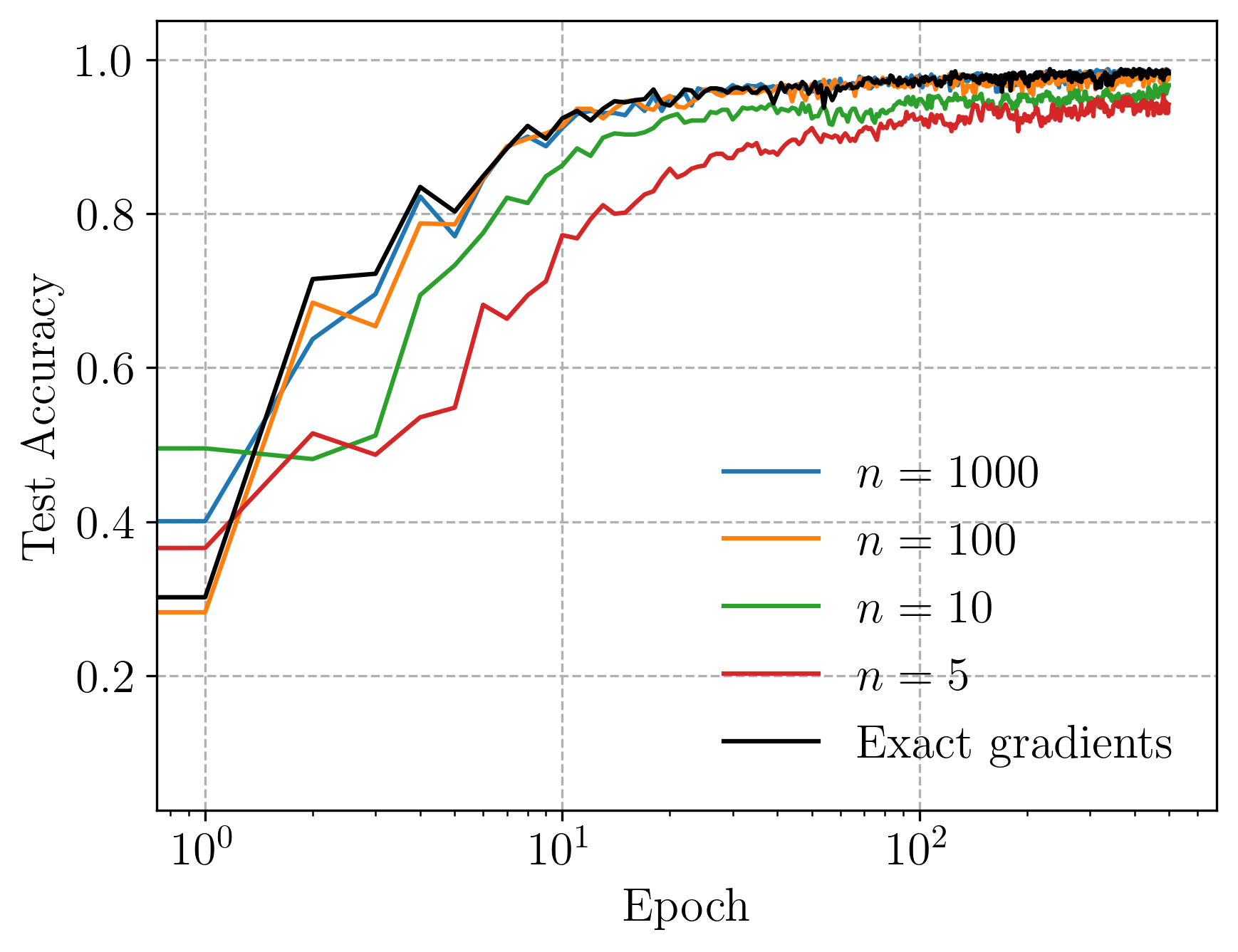}
        \caption{Learning curve for the digits dataset.}
    \label{fig:shot_noise_digits}
    \end{subfigure}
    \begin{subfigure}[b]{0.38\textwidth}
        \centering
        \includegraphics[width=\textwidth]{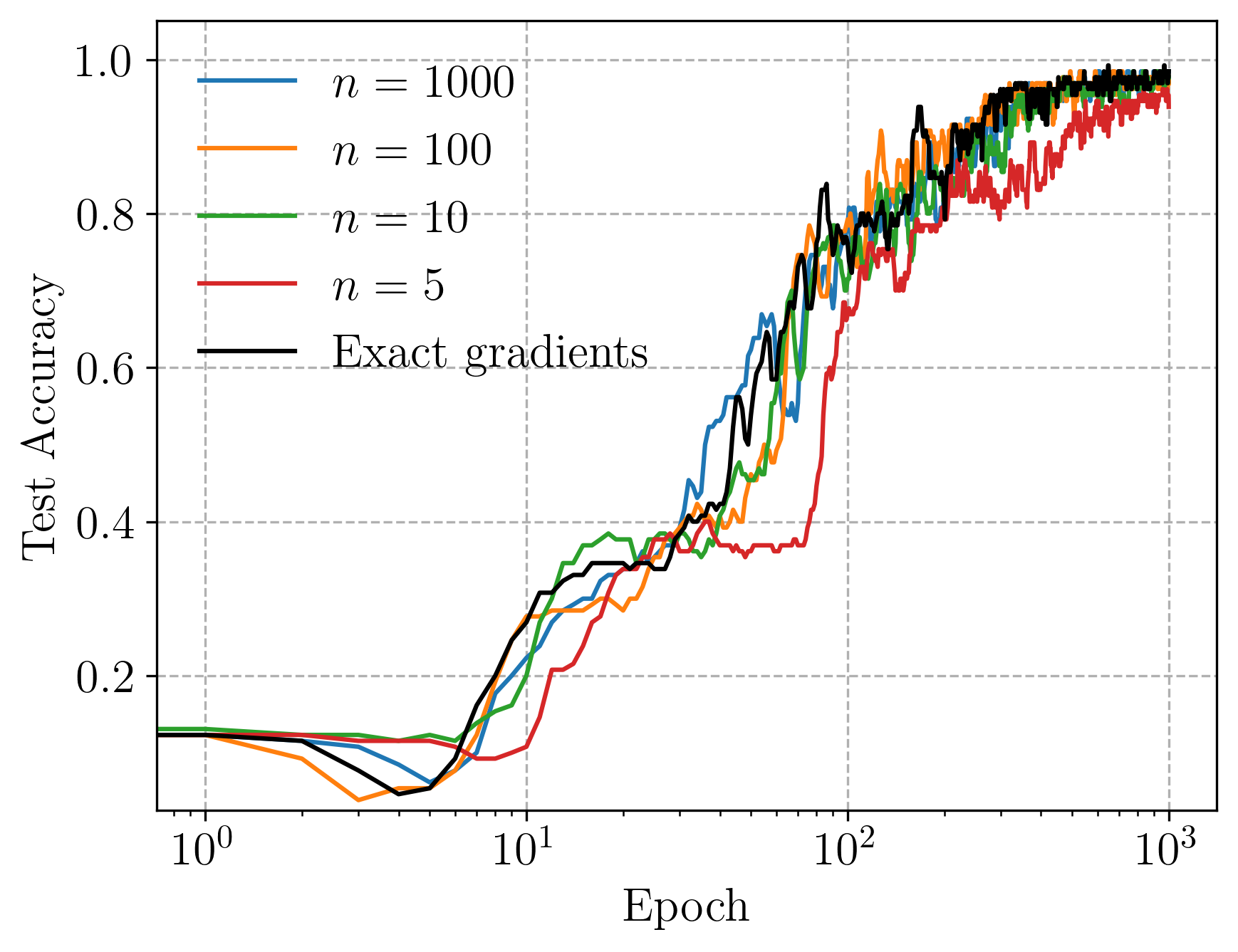}
        \caption{Learning curve for the vowel dataset.}
    \label{fig:shot_noise_vowel}
    \end{subfigure}
    \caption{Plots of model accuracy for multiclass classification problems on the test set over the course of training when using various values of $n$ to estimate gradients.}
    \label{fig:shot_noise_multiclass}
\end{figure*}

\paragraph{Hardware-based gradient determination}
We also investigated the effect that variance from parameter-shift-rule based gradient estimates have on the learning process as in the binary classification section.
The results are shown in Fig. \ref{fig:shot_noise_multiclass}.
As in the binary classification case we find that as less samples are used to estimate the gradients more epochs are typically needed for the model to converge to optimal test accuracy.
However, for the handwritten digits problem we see for $n=5$ and $n=10$ the model converges to a slightly lower test accuracy.
The same can be seen for the vowel recognition dataset when $n=5$.
For all datasets we can see that $n=100$ shots are enough for the learning curve to closely match that of the one which uses the exact gradients.

\section{Conclusion}
We have proposed a method for constructing a linear optical neural network using laser interferometry and field displacements, which is the novelty claim of this paper. 
We demonstrated that this classical optical model displays remarkable resilience to photon loss which is a prominent source of imperfections in modern photonic integrated circuits~\cite{Bogaerts2020}. 
The linear optical character of the circuit also makes it highly suitable for \textit{in situ} training using various existing training protocols based on parameter shift rules~\cite{schuld2019evaluating}, physical backpropagation~\cite{hughes2018training}, physics-aware training~\cite{wright2022deep} or forward-only training~\cite{momeni2023backpropagation, oguz2023forward, Bandyopadhyay2024, xue2024fully}.
All this makes it particularly appealing for on-chip integrated optics implementations.
We also investigated how our network performs on a variety of machine learning problems and its robustness to variance in the gradient estimates provided by parameter shift rules.

An open question is how quantum optics can be used to achieve true advantage over such linear optical circuits for classical machine learning tasks, given that previous proposals with non-Gaussian elements such as Kerr gates or photon number resolving measurements at each layer of the network have not yet outperformed linear optics. 

One straightforward quantum improvement would be to add an inline squeezer of the amplitude quadrature $Q$ before its measurement by homodyne detection during both training and inference.
The corresponding quantum noise reduction will allow proportionally fewer measurements to determine the average $q=\langle Q\rangle$ consequently enabling faster inference and gradient estimation time. 
Note that just 10 dB of squeezing reduces the variance by ten times and that on-chip quantum homodyne detection has been demonstrated at a bandwidth of 9 GHz~\cite{Tasker2021} and PIN photodiode detection bandwidth has been demonstrated to reach 265 GHz~\cite{Lischke2021}. 
Another, albeit more complex, quantum improvement is swapping the single $M \times M$ interferometer with a set of $M$ single mode squeezers which are sandwiched between two $M \times M$ interferometers.
This configuration, based on the singular value decomposition, is known to allow one to exactly mimic the functionality of fully connected linear layers within a classical neural network \cite{killoran2019continuous}.

For future work one could explore how these networks can more closely approximate the expressivity of classical neural networks. For example, classical neural networks inherently have a feed forward structure in which outputs from one layer are composed with the nonlinear transformation (activation function) in the next layer leading to the network learning hierarchical and distributed representations of the inputs at each layer~\cite{bishop2023deep}. 
An interesting example of such feedforward (though a nonlinear optical one, using field encoding rather than parameter encoding) can be found in Ref.~\cite{Bandyopadhyay2024}. 
Furthermore, the existence of universal function approximation theorems for classical neural networks \cite{cybenko1989approximation, leshno1993multilayer} provides a strong theoretical basis for their expressivity and possible extensions of these theorems to our work merits further investigation.
Lastly, extensions to implementing convolutional, recurrent and transformer layers are also important next steps. 

\begin{acknowledgments}
We thank Benjamin Scellier, Maria Schuld and Catalina Albornoz for useful discussions. OP was supported by U.S. National Science foundation grants PHY-2514971 and ECCS-2530171. 
JMS acknowledges support from the National Science Foundation via DMR-2204312. 
\end{acknowledgments}

\appendix

\section{Using intensity measurements}
\begin{figure*}[htbp]
    \centering
    \begin{subfigure}[t]{0.9\textwidth}
    \begin{minipage}[b]{0.49\linewidth}
    \centering
    \includegraphics[width=\linewidth]{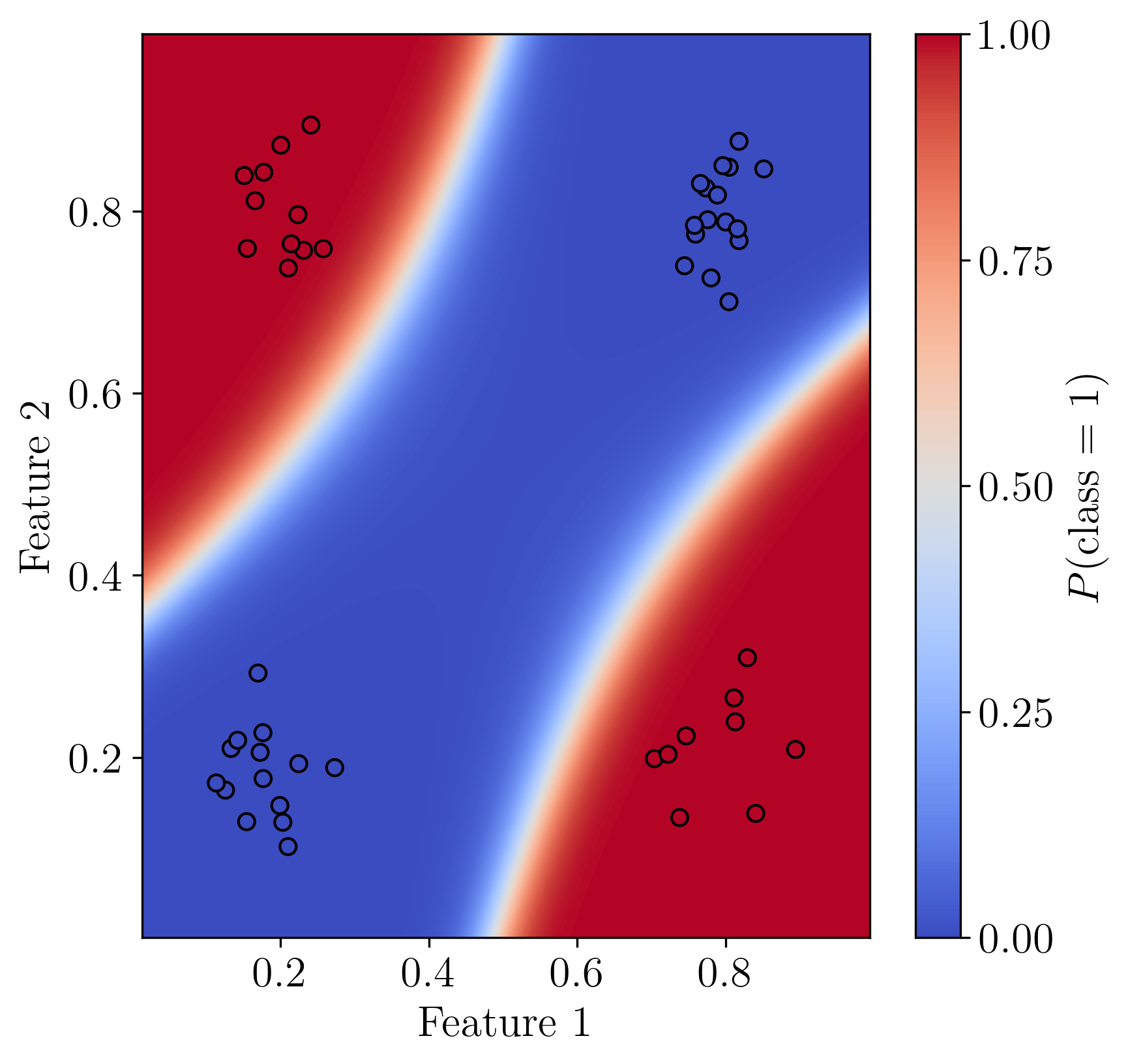}
    \end{minipage}\hfill
    \begin{minipage}[b]{0.49\linewidth}
      \centering
    \includegraphics[width=\linewidth]{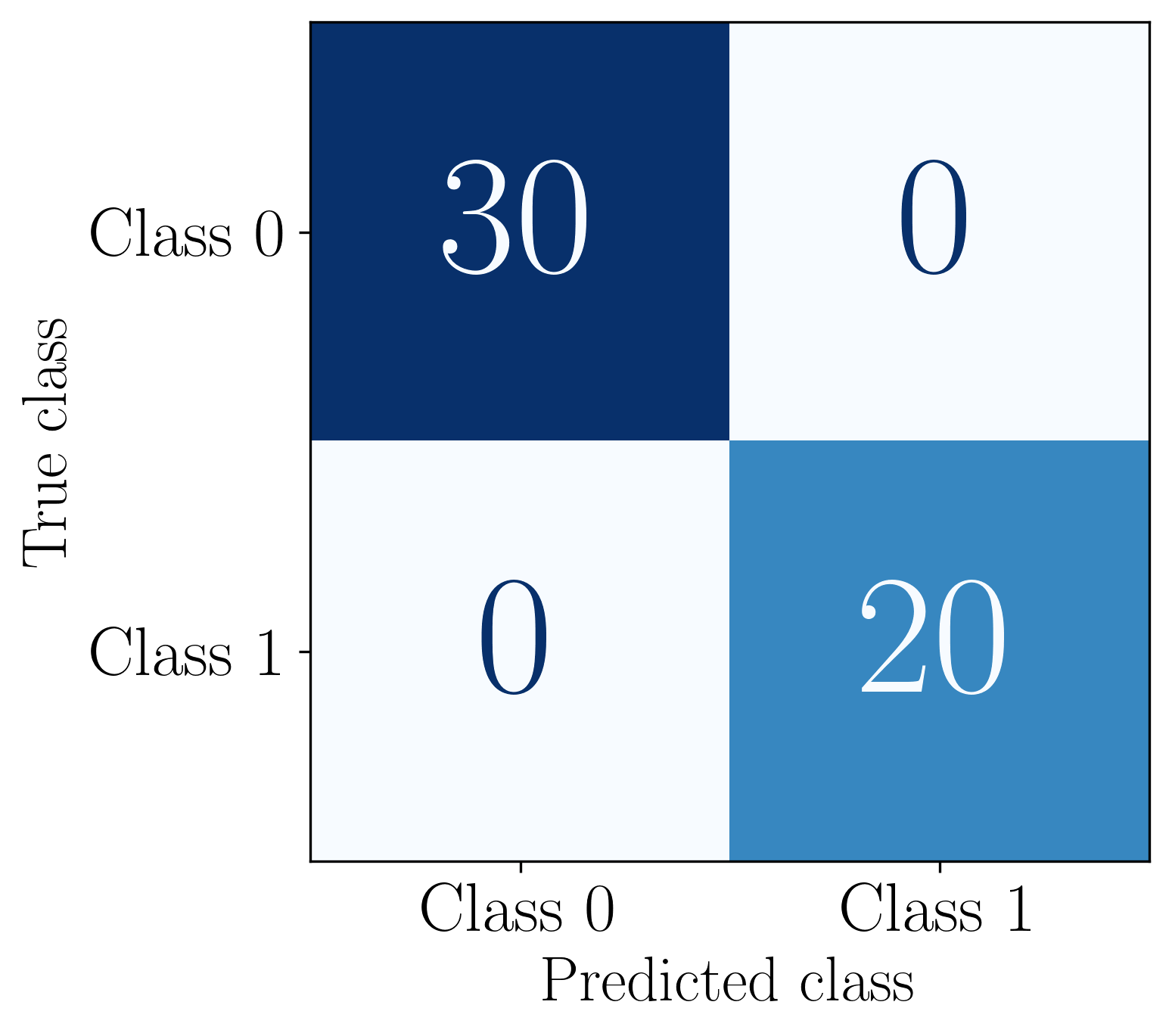}
    \end{minipage}
    \caption{Test set results for XOR classification problem.}
    \label{fig:xor}
    \end{subfigure}
    \hfill
    \begin{subfigure}[t]{0.4\textwidth}
    \centering
    \includegraphics[width=\linewidth]{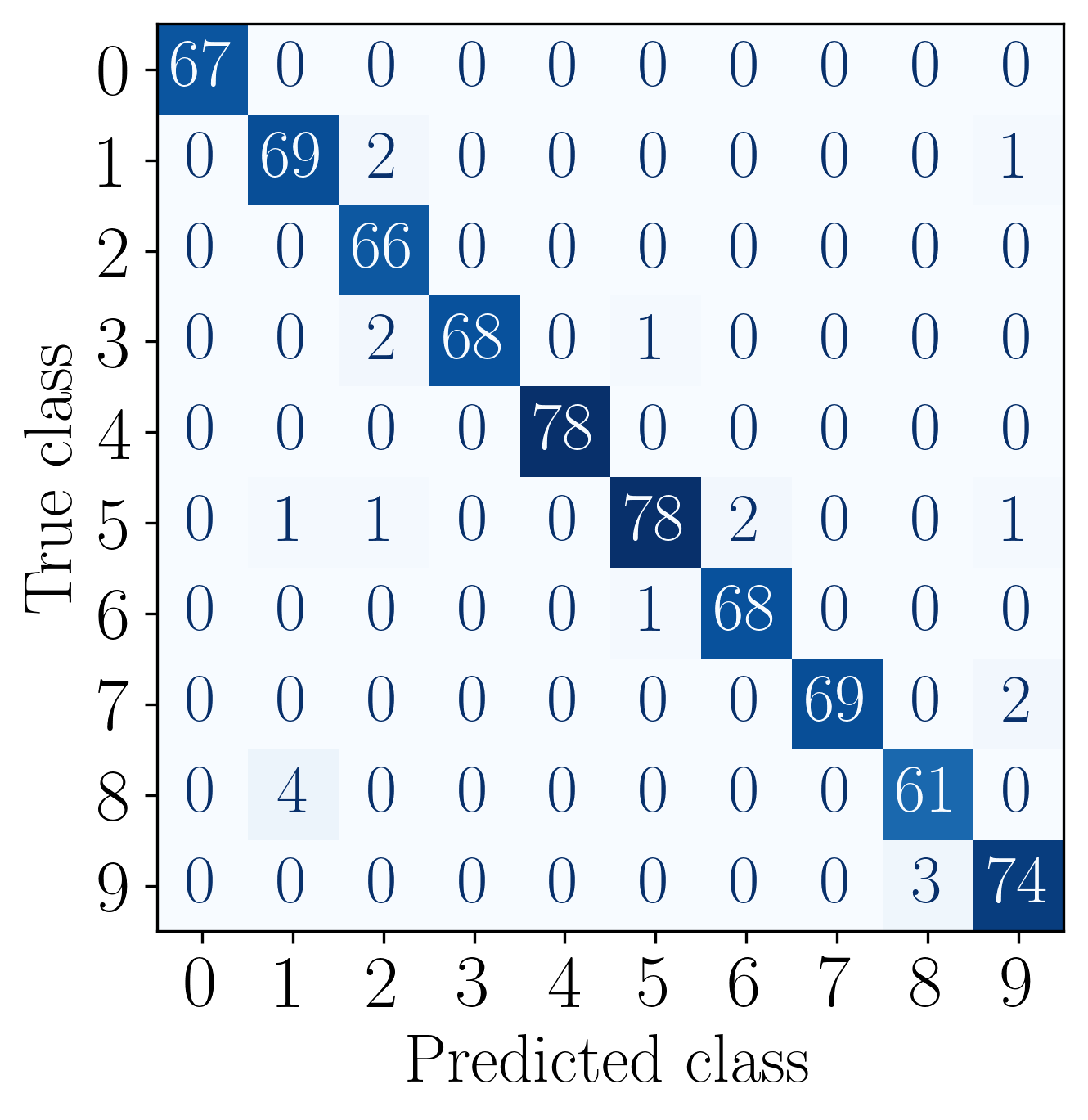}
    \caption{Test set results for handwritten digits classification problem.}
    \label{fig:digits}
    \end{subfigure}
    \hfill
    
    \caption{Classification results with ONN using intensity as output logits for the XOR and handwritten digits classification problems.}
    \label{fig:intensity_output}
\end{figure*}

Many photonic neural networks take the intensity of the optical modes to be the network output and the efficient \textit{in situ} training protocol proposed in Ref. \cite{hughes2018training} utilizes intensity measurements to extract gradients. 
Here we investigate the effect using intensity over $Q$ quadrature expectations has on our models performance.
The intensity of an optical field is defined as $\langle N \rangle$, where $N=a^\dag a$ is the photon number operator, and can be written in terms of the quadrature expectations as
\begin{align}
    \langle N \rangle = \frac{1}{2}(\langle Q^2 \rangle + \langle P^2 \rangle-1)
\end{align}
which for coherent states reduces to
\begin{align}
    \langle N \rangle = \frac{1}{2}(\langle Q \rangle^2 + \langle P \rangle^2). 
\end{align}
The results for using intensity as our network output during training and inference are shown in Fig. \ref{fig:intensity_output} for the XOR and handwritten digits classification problems using the same hyperparameters and circuit architecture considered in the main text.
For the XOR problem the test accuracy remains at 100\% while the test accuracy for the handwritten digits dataset decreases slightly to 97.08\% which we note is still comparable to that achieved by the classical neural network trained for comparison in Section \ref{sec:multilabel}.
The class probability heatmap for the XOR problem shows sharper definition of the decision boundary than the heatmap in Fig.~\ref{fig:xor_bin} of the main text. 
However, a potential disadvantage of this approach is that regression for arbitrary real valued functions becomes less straightforward as the intensity measurement result is restricted to nonnegative numbers. 

\section{Displacement regularization via the lasso}\label{disp_reg}
The additional displacements in the ONN from the main text increase energy consumption due to their active nature. 
Here we use an $L_1$ regularization strategy called LASSO regularization~\cite{james2023statistical}. 
This regularization is preferable over $L_2$ based methods like weight decay as it encourages sparsity in the penalized parameters~\cite{james2023statistical} i.e. it tends to shrink unnecessary parameters to 0. 
The regularization is implemented by adding a term to the loss function of the form
\begin{align*}
    \lambda \sum_i |\textrm{Re}(\alpha_i)| + |\textrm{Im}(\alpha_i)|
\end{align*}
where $\alpha_i$ denotes the complex displacement parameter of a particular displacement operation in the circuit and $\lambda$ denotes the strength of the regularization penalty.
Table \ref{tab:lasso_reg} shows the results. 
We can see that in many cases over half of the displacements are set to very small values of $|\alpha|$ with little to no degradation in test set performance.
This already represents a significant energy reduction without further tuning of the $\lambda$ regularization parameter. 
One may further optimize energy efficiency by simply turning off the lasers which produce displacements of small magnitude and observing the effect this has on test error. 
This is analogous to the deep learning practice of weight pruning~\cite{lecun1989optimal, hassibi1992second, han2015learning}.

\begin{table*}[ht]
\centering
\begin{tabular}{|c|c|c|}
\hline
Dataset &
\shortstack{ \% of displacements where $|\alpha|<10^{-2}$ \\ $(\lambda=0)$ } &
\shortstack{ \% of displacements where $|\alpha|<10^{-2}$ \\ $(\lambda=10^{-2})$ } \\
\hline
XOR  & $0\%$ $(100\%)$ & $44.44\%$ $(100\%)$\\
\hline
Circles  & $0\%$ $(100\%)$ & $55.56\%$ $(100\%)$\\
\hline
Moons  & $0\%$ $(100\%)$ & $66.67\%$ $(100\%)$\\
\hline
Iris Flowers  & $0\%$ $(94.67\%)$ & $83.33\%$ $(94.67\%)$\\
\hline
Vowel Recognition  & $0\%$ $(97.69\%)$ & $80.56\%$ $(97.69\%)$\\
\hline
Handwritten Digits & $0\%$ $(98.33\%)$ & $99.22\%$ $(97.77\%)$\\
\hline
\end{tabular}
\caption{Lasso regularization experiment results. The value in parenthesis denotes test accuracy. Note that for all datasets the learning rate is set to $10^{-2}$ with the remaining hyperparameters being the same as in the main text.}
\label{tab:lasso_reg}
\end{table*}

\bibliography{Anteneh, Pfister}

@article{wu2025scaling,
  title={Scaling up for end-to-end on-chip photonic neural network inference},
  author={Wu, Bo and Huang, Chaoran and Zhang, Jialong and Zhou, Hailong and Wang, Yilun and Dong, Jianji and Zhang, Xinliang},
  journal={Light: Science \& Applications},
  volume={14},
  number={1},
  pages={328},
  year={2025},
  publisher={Nature Publishing Group UK London}
}

@article{spall1998overview,
  title={An overview of the simultaneous perturbation method for efficient optimization},
  author={Spall, James C},
  journal={Johns Hopkins apl technical digest},
  volume={19},
  number={4},
  pages={482--492},
  year={1998}
}

@article{xu2026chip,
  title={On-chip input-hidden-layer-degenerate optical diffractive nonlinear neural network},
  author={Xu, Wenguang and Wu, Bo and Zhang, Shiji and Zhou, Hailong and Wang, Yilun and Dong, Jianji and Zhang, Xinliang},
  journal={Optica},
  volume={13},
  number={1},
  pages={172--179},
  year={2026},
  publisher={Optica Publishing Group}
}

@article{han2015learning,
  title={Learning both weights and connections for efficient neural network},
  author={Han, Song and Pool, Jeff and Tran, John and Dally, William},
  journal={Advances in neural information processing systems},
  volume={28},
  year={2015}
}

@article{hassibi1992second,
  title={Second order derivatives for network pruning: Optimal brain surgeon},
  author={Hassibi, Babak and Stork, David},
  journal={Advances in neural information processing systems},
  volume={5},
  year={1992}
}

@article{lecun1989optimal,
  title={Optimal brain damage},
  author={LeCun, Yann and Denker, John and Solla, Sara},
  journal={Advances in neural information processing systems},
  volume={2},
  year={1989}
}

@article{oguz2023forward,
  title={Forward--forward training of an optical neural network},
  author={Oguz, Ilker and Ke, Junjie and Weng, Qifei and Yang, Feng and Yildirim, Mustafa and Dinc, Niyazi Ulas and Hsieh, Jih-Liang and Moser, Christophe and Psaltis, Demetri},
  journal={Optics Letters},
  volume={48},
  number={20},
  pages={5249--5252},
  year={2023},
  publisher={Optica Publishing Group}
}

@article{hughes2019wave,
  title={Wave physics as an analog recurrent neural network},
  author={Hughes, Tyler W and Williamson, Ian AD and Minkov, Momchil and Fan, Shanhui},
  journal={Science advances},
  volume={5},
  number={12},
  pages={eaay6946},
  year={2019},
  publisher={American Association for the Advancement of Science}
}

@article{leshno1993multilayer,
  title={Multilayer feedforward networks with a nonpolynomial activation function can approximate any function},
  author={Leshno, Moshe and Lin, Vladimir Ya and Pinkus, Allan and Schocken, Shimon},
  journal={Neural networks},
  volume={6},
  number={6},
  pages={861--867},
  year={1993},
  publisher={Elsevier}
}

@article{anisetti2024frequency,
  title={Frequency propagation: Multimechanism learning in nonlinear physical networks},
  author={Anisetti, Vidyesh Rao and Kandala, Ananth and Scellier, Benjamin and Schwarz, JM},
  journal={Neural Computation},
  volume={36},
  number={4},
  pages={596--620},
  year={2024},
  publisher={MIT Press One Rogers Street, Cambridge, MA 02142-1209, USA journals-info~…}
}

@article{li2024training,
  title={Training all-mechanical neural networks for task learning through in situ backpropagation},
  author={Li, Shuaifeng and Mao, Xiaoming},
  journal={Nature communications},
  volume={15},
  number={1},
  pages={10528},
  year={2024},
  publisher={Nature Publishing Group UK London}
}

@article{xue2024fully,
  title={Fully forward mode training for optical neural networks},
  author={Xue, Zhiwei and Zhou, Tiankuang and Xu, Zhihao and Yu, Shaoliang and Dai, Qionghai and Fang, Lu},
  journal={Nature},
  volume={632},
  number={8024},
  pages={280--286},
  year={2024},
  publisher={Nature Publishing Group UK London}
}

@article{momeni2023backpropagation,
  title={Backpropagation-free training of deep physical neural networks},
  author={Momeni, Ali and Rahmani, Babak and Mall{\'e}jac, Matthieu and Del Hougne, Philipp and Fleury, Romain},
  journal={Science},
  volume={382},
  number={6676},
  pages={1297--1303},
  year={2023},
  publisher={American Association for the Advancement of Science}
}

@article{hillenbrand1995acoustic,
  title={Acoustic characteristics of American English vowels},
  author={Hillenbrand, James and Getty, Laura A and Clark, Michael J and Wheeler, Kimberlee},
  journal={The Journal of the Acoustical society of America},
  volume={97},
  number={5},
  pages={3099--3111},
  year={1995},
  publisher={Acoustical Society of America}
}

@article{zhang2021optical,
  title={An optical neural chip for implementing complex-valued neural network},
  author={Zhang, Hui and Gu, Mile and Jiang, XD and Thompson, Jayne and Cai, Hong and Paesani, Stefano and Santagati, Raffaele and Laing, Anthony and Zhang, Y and Yung, Man-Hong and others},
  journal={Nature communications},
  volume={12},
  number={1},
  pages={457},
  year={2021},
  publisher={Nature Publishing Group UK London}
}

@article{onodera2025arbitrary,
  title={Arbitrary control over multimode wave propagation for machine learning},
  author={Onodera, Tatsuhiro and Stein, Martin M and Ash, Benjamin A and Sohoni, Mandar M and Bosch, Melissa and Yanagimoto, Ryotatsu and Jankowski, Marc and McKenna, Timothy P and Wang, Tianyu and Shvets, Gennady and others},
  journal={Nature Physics},
  pages={1--8},
  year={2025},
  publisher={Nature Publishing Group UK London}
}

@article{romera2018vowel,
  title={Vowel recognition with four coupled spin-torque nano-oscillators},
  author={Romera, Miguel and Talatchian, Philippe and Tsunegi, Sumito and Abreu Araujo, Flavio and Cros, Vincent and Bortolotti, Paolo and Trastoy, Juan and Yakushiji, Kay and Fukushima, Akio and Kubota, Hitoshi and others},
  journal={Nature},
  volume={563},
  number={7730},
  pages={230--234},
  year={2018},
  publisher={Nature Publishing Group UK London}
}

@article{wright2022deep,
  title={Deep physical neural networks trained with backpropagation},
  author={Wright, Logan G and Onodera, Tatsuhiro and Stein, Martin M and Wang, Tianyu and Schachter, Darren T and Hu, Zoey and McMahon, Peter L},
  journal={Nature},
  volume={601},
  number={7894},
  pages={549--555},
  year={2022},
  publisher={Nature Publishing Group UK London}
}

@article{williamson2019reprogrammable,
  title={Reprogrammable electro-optic nonlinear activation functions for optical neural networks},
  author={Williamson, Ian AD and Hughes, Tyler W and Minkov, Momchil and Bartlett, Ben and Pai, Sunil and Fan, Shanhui},
  journal={IEEE Journal of Selected Topics in Quantum Electronics},
  volume={26},
  number={1},
  pages={1--12},
  year={2019},
  publisher={IEEE}
}

@article{lopez2023self,
  title={Self-learning machines based on Hamiltonian echo backpropagation},
  author={Lopez-Pastor, Victor and Marquardt, Florian},
  journal={Physical Review X},
  volume={13},
  number={3},
  pages={031020},
  year={2023},
  publisher={APS}
}

@article{hughes2018training,
  title={Training of photonic neural networks through in situ backpropagation and gradient measurement},
  author={Hughes, Tyler W and Minkov, Momchil and Shi, Yu and Fan, Shanhui},
  journal={Optica},
  volume={5},
  number={7},
  pages={864--871},
  year={2018},
  publisher={Optical Society of America}
}

@article{shainline2017superconducting,
  title={Superconducting optoelectronic circuits for neuromorphic computing},
  author={Shainline, Jeffrey M and Buckley, Sonia M and Mirin, Richard P and Nam, Sae Woo},
  journal={Physical Review Applied},
  volume={7},
  number={3},
  pages={034013},
  year={2017},
  publisher={APS}
}

@article{ashtiani2022chip,
  title={An on-chip photonic deep neural network for image classification},
  author={Ashtiani, Farshid and Geers, Alexander J and Aflatouni, Firooz},
  journal={Nature},
  volume={606},
  number={7914},
  pages={501--506},
  year={2022},
  publisher={Nature Publishing Group UK London}
}

@article{shen2017deep,
  title={Deep learning with coherent nanophotonic circuits},
  author={Shen, Yichen and Harris, Nicholas C and Skirlo, Scott and Prabhu, Mihika and Baehr-Jones, Tom and Hochberg, Michael and Sun, Xin and Zhao, Shijie and Larochelle, Hugo and Englund, Dirk and others},
  journal={Nature photonics},
  volume={11},
  number={7},
  pages={441--446},
  year={2017},
  publisher={Nature Publishing Group UK London}
}

@article{cin2025training,
  title={Training nonlinear optical neural networks with Scattering Backpropagation},
  author={Cin, Nicola Dal and Marquardt, Florian and Wanjura, Clara C},
  journal={arXiv preprint arXiv:2508.11750},
  year={2025}
}

@inproceedings{jing2017tunable,
  title={Tunable efficient unitary neural networks (eunn) and their application to rnns},
  author={Jing, Li and Shen, Yichen and Dubcek, Tena and Peurifoy, John and Skirlo, Scott and LeCun, Yann and Tegmark, Max and Solja{\v{c}}i{\'c}, Marin},
  booktitle={International Conference on Machine Learning},
  pages={1733--1741},
  year={2017},
  organization={PMLR}
}

@article{chen2023deep,
  title={Deep learning with coherent VCSEL neural networks},
  author={Chen, Zaijun and Sludds, Alexander and Davis III, Ronald and Christen, Ian and Bernstein, Liane and Ateshian, Lamia and Heuser, Tobias and Heermeier, Niels and Lott, James A and Reitzenstein, Stephan and others},
  journal={Nature Photonics},
  volume={17},
  number={8},
  pages={723--730},
  year={2023},
  publisher={Nature Publishing Group UK London}
}

@article{hamerly2019large,
  title={Large-scale optical neural networks based on photoelectric multiplication},
  author={Hamerly, Ryan and Bernstein, Liane and Sludds, Alexander and Solja{\v{c}}i{\'c}, Marin and Englund, Dirk},
  journal={Physical Review X},
  volume={9},
  number={2},
  pages={021032},
  year={2019},
  publisher={APS}
}

@book{lemons2002introduction,
  title={An introduction to stochastic processes in physics},
  author={Lemons, Don S and Langevin, Paul},
  year={2002},
  publisher={JHU Press}
}

@article{reck1994experimental,
  title={Experimental realization of any discrete unitary operator},
  author={Reck, Michael and Zeilinger, Anton and Bernstein, Herbert J and Bertani, Philip},
  journal={Physical review letters},
  volume={73},
  number={1},
  pages={58},
  year={1994},
  publisher={APS}
}

@article{wang2022optical,
  title={An optical neural network using less than 1 photon per multiplication},
  author={Wang, Tianyu and Ma, Shi-Yuan and Wright, Logan G and Onodera, Tatsuhiro and Richard, Brian C and McMahon, Peter L},
  journal={Nature Communications},
  volume={13},
  number={1},
  pages={123},
  year={2022},
  publisher={Nature Publishing Group UK London}
}

@article{wang2024training,
  title={Training coupled phase oscillators as a neuromorphic platform using equilibrium propagation},
  author={Wang, Qingshan and Wanjura, Clara C and Marquardt, Florian},
  journal={Neuromorphic Computing and Engineering},
  volume={4},
  number={3},
  pages={034014},
  year={2024},
  publisher={IOP Publishing}
}

@book{cerf2007quantum,
  title={Quantum information with continuous variables of atoms and light},
  author={Cerf, Nicolas J and Leuchs, Gerd and Polzik, Eugene S},
  year={2007},
  publisher={World Scientific}
}

@article{austin2025hybrid,
  title={Hybrid quantum-classical photonic neural networks},
  author={Austin, Tristan and Bilodeau, Simon and Hayman, Andrew and Rotenberg, Nir and Shastri, Bhavin J},
  journal={npj Unconventional Computing},
  volume={2},
  number={1},
  pages={29},
  year={2025},
  publisher={Nature Publishing Group UK London}
}

@article{xia2024nonlinear,
  title={Nonlinear optical encoding enabled by recurrent linear scattering},
  author={Xia, Fei and Kim, Kyungduk and Eliezer, Yaniv and Han, SeungYun and Shaughnessy, Liam and Gigan, Sylvain and Cao, Hui},
  journal={Nature Photonics},
  volume={18},
  number={10},
  pages={1067--1075},
  year={2024},
  publisher={Nature Publishing Group UK London}
}

@article{yildirim2024nonlinear,
  title={Nonlinear processing with linear optics},
  author={Yildirim, Mustafa and Dinc, Niyazi Ulas and Oguz, Ilker and Psaltis, Demetri and Moser, Christophe},
  journal={Nature Photonics},
  volume={18},
  number={10},
  pages={1076--1082},
  year={2024},
  publisher={Nature Publishing Group UK London}
}

@article{fisher1936use,
  title={The use of multiple measurements in taxonomic problems},
  author={Fisher, Ronald A},
  journal={Annals of eugenics},
  volume={7},
  number={2},
  pages={179--188},
  year={1936},
  publisher={Wiley Online Library}
}

@article{ji2025photonic,
  title={Photonic neuromorphic computing using symmetry-protected zero modes in coupled nanolaser arrays},
  author={Ji, Kaiwen and Tirabassi, Giulio and Masoller, Cristina and Ge, Li and Yacomotti, Alejandro M},
  journal={Nature communications},
  volume={16},
  number={1},
  pages={9203},
  year={2025},
  publisher={Nature Publishing Group UK London}
}

@article{wanjura2025quantum,
  title={Quantum equilibrium propagation for efficient training of quantum systems based on Onsager reciprocity},
  author={Wanjura, Clara C and Marquardt, Florian},
  journal={Nature Communications},
  volume={16},
  number={1},
  pages={6595},
  year={2025},
  publisher={Nature Publishing Group UK London}
}

@article{bergholm2018pennylane,
  title={Pennylane: Automatic differentiation of hybrid quantum-classical computations},
  author={Bergholm, Ville and Izaac, Josh and Schuld, Maria and Gogolin, Christian and Ahmed, Shahnawaz and Ajith, Vishnu and Alam, M Sohaib and Alonso-Linaje, Guillermo and AkashNarayanan, B and Asadi, Ali and others},
  journal={arXiv preprint arXiv:1811.04968},
  year={2018}
}

@article{pedregosa2011scikit,
  title={Scikit-learn: Machine learning in Python},
  author={Pedregosa, Fabian and Varoquaux, Ga{\"e}l and Gramfort, Alexandre and Michel, Vincent and Thirion, Bertrand and Grisel, Olivier and Blondel, Mathieu and Prettenhofer, Peter and Weiss, Ron and Dubourg, Vincent and others},
  journal={the Journal of machine Learning research},
  volume={12},
  pages={2825--2830},
  year={2011},
  publisher={JMLR. org}
}

@article{anisetti2023learning,
  title={Learning by non-interfering feedback chemical signaling in physical networks},
  author={Anisetti, Vidyesh Rao and Scellier, Benjamin and Schwarz, Jennifer M},
  journal={Physical Review Research},
  volume={5},
  number={2},
  pages={023024},
  year={2023},
  publisher={APS}
}

@article{kubler2020adaptive,
  title={An adaptive optimizer for measurement-frugal variational algorithms},
  author={K{\"u}bler, Jonas M and Arrasmith, Andrew and Cincio, Lukasz and Coles, Patrick J},
  journal={Quantum},
  volume={4},
  pages={263},
  year={2020},
  publisher={Verein zur F{\"o}rderung des Open Access Publizierens in den Quantenwissenschaften}
}

@article{adesso2014continuous,
  title={Continuous variable quantum information: Gaussian states and beyond},
  author={Adesso, Gerardo and Ragy, Sammy and Lee, Antony R},
  journal={Open Systems \& Information Dynamics},
  volume={21},
  number={01n02},
  pages={1440001},
  year={2014},
  publisher={World Scientific}
}

@article{scellier2024quantum,
  title={Quantum equilibrium propagation: Gradient-descent training of quantum systems},
  author={Scellier, Benjamin},
  journal={arXiv preprint arXiv:2406.00879},
  year={2024}
}

@article{scellier2017equilibrium,
  title={Equilibrium propagation: Bridging the gap between energy-based models and backpropagation},
  author={Scellier, Benjamin and Bengio, Yoshua},
  journal={Frontiers in computational neuroscience},
  volume={11},
  pages={24},
  year={2017},
  publisher={Frontiers Media SA}
}

@article{sweke2020stochastic,
  title={Stochastic gradient descent for hybrid quantum-classical optimization},
  author={Sweke, Ryan and Wilde, Frederik and Meyer, Johannes and Schuld, Maria and F{\"a}hrmann, Paul K and Meynard-Piganeau, Barth{\'e}l{\'e}my and Eisert, Jens},
  journal={Quantum},
  volume={4},
  pages={314},
  year={2020},
  publisher={Verein zur F{\"o}rderung des Open Access Publizierens in den Quantenwissenschaften}
}

@article{bangar2023experimentally,
  title={Experimentally realizable continuous-variable quantum neural networks},
  author={Bangar, Shikha and Sunny, Leanto and Yeter-Aydeniz, K{\"u}bra and Siopsis, George},
  journal={Physical Review A},
  volume={108},
  number={4},
  pages={042414},
  year={2023},
  publisher={APS}
}

@article{killoran2019continuous,
  title={Continuous-variable quantum neural networks},
  author={Killoran, Nathan and Bromley, Thomas R and Arrazola, Juan Miguel and Schuld, Maria and Quesada, Nicol{\'a}s and Lloyd, Seth},
  journal={Physical Review Research},
  volume={1},
  number={3},
  pages={033063},
  year={2019},
  publisher={APS}
}

@article{pai2023experimentally,
  title={Experimentally realized in situ backpropagation for deep learning in photonic neural networks},
  author={Pai, Sunil and Sun, Zhanghao and Hughes, Tyler W and Park, Taewon and Bartlett, Ben and Williamson, Ian AD and Minkov, Momchil and Milanizadeh, Maziyar and Abebe, Nathnael and Morichetti, Francesco and others},
  journal={Science},
  volume={380},
  number={6643},
  pages={398--404},
  year={2023},
  publisher={American Association for the Advancement of Science}
}

@article{markovic2020physics,
  title={Physics for neuromorphic computing},
  author={Markovi{\'c}, Danijela and Mizrahi, Alice and Querlioz, Damien and Grollier, Julie},
  journal={Nature Reviews Physics},
  volume={2},
  number={9},
  pages={499--510},
  year={2020},
  publisher={Nature Publishing Group UK London}
}

@article{christensen20222022,
  title={2022 roadmap on neuromorphic computing and engineering},
  author={Christensen, Dennis V and Dittmann, Regina and Linares-Barranco, Bernabe and Sebastian, Abu and Le Gallo, Manuel and Redaelli, Andrea and Slesazeck, Stefan and Mikolajick, Thomas and Spiga, Sabina and Menzel, Stephan and others},
  journal={Neuromorphic Computing and Engineering},
  volume={2},
  number={2},
  pages={022501},
  year={2022},
  publisher={IOP Publishing}
}

@article{schuld2019evaluating,
  title={Evaluating analytic gradients on quantum hardware},
  author={Schuld, Maria and Bergholm, Ville and Gogolin, Christian and Izaac, Josh and Killoran, Nathan},
  journal={Physical Review A},
  volume={99},
  number={3},
  pages={032331},
  year={2019},
  publisher={APS}
}

@article{momeni2025training,
  title={Training of physical neural networks},
  author={Momeni, Ali and Rahmani, Babak and Scellier, Benjamin and Wright, Logan G and McMahon, Peter L and Wanjura, Clara C and Li, Yuhang and Skalli, Anas and Berloff, Natalia G and Onodera, Tatsuhiro and others},
  journal={Nature},
  volume={645},
  number={8079},
  pages={53--61},
  year={2025},
  publisher={Nature Publishing Group UK London}
}

@article{wanjura2024fully,
  title={Fully nonlinear neuromorphic computing with linear wave scattering},
  author={Wanjura, Clara C and Marquardt, Florian},
  journal={Nature Physics},
  volume={20},
  number={9},
  pages={1434--1440},
  year={2024},
  publisher={Nature Publishing Group UK London}
}

@book{goodfellow2016deep,
  title={Deep learning},
  author={Goodfellow, Ian and Bengio, Yoshua and Courville, Aaron and Bengio, Yoshua},
  volume={1},
  number={2},
  year={2016},
  publisher={MIT press Cambridge}
}

@book{bishop2023deep,
  title={Deep learning: Foundations and concepts},
  author={Bishop, Christopher M and Bishop, Hugh},
  year={2023},
  publisher={Springer Nature}
}

@article{cybenko1989approximation,
  title={Approximation by superpositions of a sigmoidal function},
  author={Cybenko, George},
  journal={Mathematics of control, signals and systems},
  volume={2},
  number={4},
  pages={303--314},
  year={1989},
  publisher={Springer}
}

@article{paszke2019pytorch,
  title={Pytorch: An imperative style, high-performance deep learning library},
  author={Paszke, Adam and Gross, Sam and Massa, Francisco and Lerer, Adam and Bradbury, James and Chanan, Gregory and Killeen, Trevor and Lin, Zeming and Gimelshein, Natalia and Antiga, Luca and others},
  journal={Advances in neural information processing systems},
  volume={32},
  year={2019}
}

@book{james2023statistical,
  title={An introduction to statistical learning: With applications in Python},
  author={James, Gareth and Witten, Daniela and Hastie, Trevor and Tibshirani, Robert and Taylor, Jonathan},
  booktitle={An introduction to statistical learning: With applications in Python},
  year={2023},
  publisher={Springer}
}

@book{devore2021modern,
  title={Modern mathematical statistics with applications},
  author={Devore, Jay L and Berk, Kenneth N and Carlton, Matthew A},
  year={2021},
  publisher={Springer Nature}
}

@article{raffel2020exploring,
  title={Exploring the limits of transfer learning with a unified text-to-text transformer},
  author={Raffel, Colin and Shazeer, Noam and Roberts, Adam and Lee, Katherine and Narang, Sharan and Matena, Michael and Zhou, Yanqi and Li, Wei and Liu, Peter J},
  journal={Journal of machine learning research},
  volume={21},
  number={140},
  pages={1--67},
  year={2020}
}

@article{brown2020language,
  title={Language models are few-shot learners},
  author={Brown, Tom and Mann, Benjamin and Ryder, Nick and Subbiah, Melanie and Kaplan, Jared D and Dhariwal, Prafulla and Neelakantan, Arvind and Shyam, Pranav and Sastry, Girish and Askell, Amanda and others},
  journal={Advances in neural information processing systems},
  volume={33},
  pages={1877--1901},
  year={2020}
}

@article{touvron2023llama,
  title={Llama: Open and efficient foundation language models},
  author={Touvron, Hugo and Lavril, Thibaut and Izacard, Gautier and Martinet, Xavier and Lachaux, Marie-Anne and Lacroix, Timoth{\'e}e and Rozi{\`e}re, Baptiste and Goyal, Naman and Hambro, Eric and Azhar, Faisal and others},
  journal={arXiv preprint arXiv:2302.13971},
  year={2023}
}

@inproceedings{kingma2015adam,
  author    = {Kingma, Diederik P. and Ba, Jimmy},
  title     = {Adam: A Method for Stochastic Optimization},
  booktitle = {International Conference on Learning Representations (ICLR)},
  year      = {2015}
}

@string{apl = {App. Phys. Lett.}}

@article{Lischke2021,
	author = {Lischke, S. and Peczek, A. and Morgan, J. S. and Sun, K. and Steckler, D. and Yamamoto, Y. and Kornd{\"o}rfer, F. and Mai, C. and Marschmeyer, S. and Fraschke, M. and Kr{\"u}ger, A. and Beling, A. and Zimmermann, L.},
	date = {2021/12/01},
	doi = {10.1038/s41566-021-00893-w},
	id = {Lischke2021},
	isbn = {1749-4893},
	journal = {Nature Photonics},
	number = {12},
	pages = {925--931},
	title = {Ultra-fast germanium photodiode with 3-dB bandwidth of 265 GHz},
	url = {https://doi.org/10.1038/s41566-021-00893-w},
	volume = {15},
	year = {2021}}

@article{Tasker2021,
	author = {Tasker, Joel F. and Frazer, Jonathan and Ferranti, Giacomo and Allen, Euan J. and Brunel, L{\'e}andre F. and Tanzilli, S{\'e}bastien and D'Auria, Virginia and Matthews, Jonathan C. F.},
	date = {2021/01/01},
	doi = {10.1038/s41566-020-00715-5},
	id = {Tasker2021},
	isbn = {1749-4893},
	journal = {Nature Photonics},
	number = {1},
	pages = {11--15},
	title = {Silicon photonics interfaced with integrated electronics for 9 GHz measurement of squeezed light},
	url = {https://doi.org/10.1038/s41566-020-00715-5},
	volume = {15},
	year = {2021}}

@article{Bandyopadhyay2024,
	author = {Bandyopadhyay, Saumil and Sludds, Alexander and Krastanov, Stefan and Hamerly, Ryan and Harris, Nicholas and Bunandar, Darius and Streshinsky, Matthew and Hochberg, Michael and Englund, Dirk},
	date = {2024/12/01},
	doi = {10.1038/s41566-024-01567-z},
	id = {Bandyopadhyay2024},
	isbn = {1749-4893},
	journal = {Nature Photonics},
	number = {12},
	pages = {1335--1343},
	title = {Single-chip photonic deep neural network with forward-only training},
	url = {https://doi.org/10.1038/s41566-024-01567-z},
	volume = {18},
	year = {2024}}

@article{Bogaerts2020,
	author = {Bogaerts, Wim and P{\'e}rez, Daniel and Capmany, Jos{\'e} and Miller, David A. B. and Poon, Joyce and Englund, Dirk and Morichetti, Francesco and Melloni, Andrea},
	date = {2020/10/01},
	doi = {10.1038/s41586-020-2764-0},
	id = {Bogaerts2020},
	isbn = {1476-4687},
	journal = {Nature},
	number = {7828},
	pages = {207--216},
	title = {Programmable photonic circuits},
	url = {https://doi.org/10.1038/s41586-020-2764-0},
	volume = {586},
	year = {2020}}

@article{PT2024,
	author = {Johanna L. Miller},
	doi = {10.1063/pt.vbbo.lurd},
	journal = {Physics Today},
	number = {10},
	pages = {12--14},
	publisher = {AIP Publishing LLC},
	title = {Nonlinear optical computing doesn't need nonlinear optics},
	volume = {77},
	year = {2024}}

@article{Clements2016,
	author = {William R. Clements and Peter C. Humphreys and Benjamin J. Metcalf and W. Steven Kolthammer and Ian A. Walmsley},
	doi = {10.1364/OPTICA.3.001460},
	journal = {Optica},
	month = {Dec},
	number = {12},
	pages = {1460--1465},
	publisher = {Optica Publishing Group},
	title = {Optimal design for universal multiport interferometers},
	url = {https://opg.optica.org/optica/abstract.cfm?URI=optica-3-12-1460},
	volume = {3},
	year = {2016}}

@article{Pfister2019,
	author = {Olivier Pfister},
	doi = {10.1088/1361-6455/ab526f},
	journal = {Journal of Physics B: Atomic, Molecular and Optical Physics},
	month = {nov},
	number = {1},
	pages = {012001},
	publisher = {{IOP} Publishing},
	title = {Continuous-variable quantum computing in the quantum optical frequency comb},
	url = {https://doi.org/10.1088/1361-6455/ab526f},
	volume = {53},
	year = 2020}

@article{Asavanant2019,
	author = {Asavanant, Warit and Shiozawa, Yu and Yokoyama, Shota and Charoensombutamon, Baramee and Emura, Hiroki and Alexander, Rafael N. and Takeda, Shuntaro and Yoshikawa, Jun-ichi and Menicucci, Nicolas C. and Yonezawa, Hidehiro and Furusawa, Akira},
	doi = {10.1126/science.aay2645},
	eprint = {https://science.sciencemag.org/content/366/6463/373.full.pdf},
	issn = {0036-8075},
	journal = {Science},
	number = {6463},
	pages = {373--376},
	publisher = {American Association for the Advancement of Science},
	title = {Generation of time-domain-multiplexed two-dimensional cluster state},
	url = {https://science.sciencemag.org/content/366/6463/373},
	volume = {366},
	year = {2019}}

@article{Larsen2019,
	author = {Larsen, Mikkel V. and Guo, Xueshi and Breum, Casper R. and Neergaard-Nielsen, Jonas S. and Andersen, Ulrik L.},
	doi = {10.1126/science.aay4354},
	eprint = {https://science.sciencemag.org/content/366/6463/369.full.pdf},
	issn = {0036-8075},
	journal = {Science},
	number = {6463},
	pages = {369--372},
	publisher = {American Association for the Advancement of Science},
	title = {Deterministic generation of a two-dimensional cluster state},
	url = {https://science.sciencemag.org/content/366/6463/369},
	volume = {366},
	year = {2019}}

@article{Yoshikawa2016,
	author = {Yoshikawa, Jun-ichi and Yokoyama, Shota and Kaji, Toshiyuki and Sornphiphatphong, Chanond and Shiozawa, Yu and Makino, Kenzo and Furusawa, Akira},
	doi = {http://dx.doi.org/10.1063/1.4962732},
	eid = 060801,
	journal = {APL Photonics},
	number = {6},
	pages = {060801},
	title = {Invited Article: Generation of one-million-mode continuous-variable cluster state by unlimited time-domain multiplexing},
	url = {http://scitation.aip.org/content/aip/journal/app/1/6/10.1063/1.4962732},
	volume = {1},
	year = {2016}}

@article{Chen2014,
	author = {Chen, Moran and Menicucci, Nicolas C. and Pfister, Olivier},
	doi = {10.1103/PhysRevLett.112.120505},
	issue = {12},
	journal = {Phys. Rev. Lett.},
	month = {Mar},
	numpages = {5},
	pages = {120505},
	publisher = {American Physical Society},
	title = {Experimental realization of multipartite entanglement of 60 modes of a quantum optical frequency comb},
	url = {http://link.aps.org/doi/10.1103/PhysRevLett.112.120505},
	volume = {112},
	year = {2014}}

@article{Weedbrook2012,
	author = {Weedbrook, Christian and Pirandola, Stefano and Garc\'{\i}a-Patr\'on, Ra\'ul and Cerf, Nicolas J. and Ralph, Timothy C. and Shapiro, Jeffrey H. and Lloyd, Seth},
	doi = {10.1103/RevModPhys.84.621},
	issue = {2},
	journal = {Rev. Mod. Phys.},
	month = {May},
	pages = {621--669},
	publisher = {American Physical Society},
	title = {Gaussian quantum information},
	url = {http://link.aps.org/doi/10.1103/RevModPhys.84.621},
	volume = {84},
	year = {2012}}

@article{Braunstein2005a,
	author = {S. L. Braunstein and P. van Loock},
	doi = {10.1103/RevModPhys.77.513},
	eid = {513},
	journal = {Rev. Mod. Phys.},
	number = {2},
	pages = {513},
	publisher = {APS},
	title = {Quantum information with continuous variables},
	volume = {77},
	web-url = {http://link.aps.org/abstract/RMP/v77/p513},
	year = {2005}}

\end{document}